\documentclass[]{aa}

\usepackage{graphicx}
\usepackage{txfonts}
\begin{document}

   \title{Triple-frequency meteor radar full wave scattering}

   \subtitle{Measurements and comparison to theory}

   \author{G. Stober
          \inst{1,2} \and
          P. Brown\inst{2,3}
          \and
         M. Campbell-Brown\inst{2,3}
         \and
          R.J. Weryk\inst{4}}

   \institute {University of Bern \& Oeschger Center for Climate Change Research, Microwave Physics, Bern, Switzerland
              \email{gunter.stober@iap.unibe.ch}
              \and
            University of Western Ontario, Department of Physics and Astronomy, London, Ontario, N6A 3K7, Canada\\
             \email{pbrown@uwo.ca}
              \and
              Institute for Earth and Space Exploration, University of Western Ontario, London, Ontario, Canada N6A 5B7
               \and
             Institute for Astronomy, University of Hawaii, 2680 Woodlawn Drive, Honolulu, HI, 96822, USA
             \email{weryk@hawaii.edu}
                         }

   \date{}


  \abstract
   {Radar scattering from meteor trails depends on several poorly constrained quantities, such as electron line density, $q$, initial trail radius, $r_0$, and ambipolar diffusion coefficient, D.}
   {The goal is to apply a numerical model of full wave backscatter to triple frequency echo measurements to validate theory and constrain estimates of electron radial distribution, initial trail radius, and the ambipolar diffusion coefficient.}
   {A selection of 50 transversely polarized and 50 parallel polarized echoes with complete trajectory information were identified from simultaneous tri-frequency echoes recorded by the Canadian Meteor Orbit Radar (CMOR). The amplitude-time profile of each echo was fit to our model using three different choices for the radial electron distribution assuming a Gaussian, parabolic-exponential, and 1-by-r$^2$ electron line density model. The observations were manually fit by varying, $q$, $r_0$, and D per model until all three synthetic echo-amplitude profiles at each frequency matched observation. }
   {The Gaussian radial electron distribution was the most successful at fitting echo power profiles, followed by the $1/r^2$. We were unable to fit any echoes using a profile where electron density varied from the trail axis as an exponential-parabolic distribution. While fewer than 5\% of all examined echoes had self-consistent fits, the estimates of $r_0$ and D as a function of height obtained were broadly similar to earlier studies, though with considerable scatter. Most meteor echoes are found to not be described well by the idealized full wave scattering model.}
   {}


   \maketitle
%

\section{Introduction}

Transverse radio wave scattering from meteor trails (also called specular reflections) have been used to infer properties of the middle atmosphere and astronomical information related to the meteoroid environment of the Solar System for 70 years \citep{Baggaley2009}. The basic physics of electromagnetic scattering from a long cylindrical trail of electrons and ions has been understood in broad form since the late 1940s \citep{lovell1948, Herlofson_1951, Kaiser1952, POULTER_and_Baggaley_1977_full_wave_scat, Ceplecha1998}.

In addition to their original use in meteor astronomy for such questions as the meteoroid velocity distribution and meteoroid mass influx  \citep{Hocking_2000_meteor_speed_measure,Elford_2004_Fresnel_holography,Holdsworth_2007,STOBER_2011_meteoroid_mass_velocity,Mazur2020}, transverse scatter radars have become standard observation systems for mesosphere and lower thermosphere dynamics. These aeronomy measurements provide a useful benchmark for the validation of general circulation models \citep{McCormack_2017_MLT_winds} at these altitudes including estimates of temperature and winds.

In recent years, meteor trail modeling efforts have focused on two distinct types of meteor echoes: meteor head echoes and non-specular echoes. Meteor head echoes are radial reflections from the dense plasma co-moving with a meteoroid \citep{Close_et_al_head_echo_scattering, Marshall_and_Close_2015_FTDT, Sugar:2018_plasma_model_theory_and_simulation} and reflect scattering on scales comparable to the mean free path and timescales on the order of milliseconds. The other type of echo, non-specular echoes \citep{Dyrud2007}, are long-lived reflections driven either by plasma instabilities related to the orientation of the local geomagnetic field, rather than the orientation of the trail alone \citep{Oppenheim2003, close2008}, or by the presence of charged dust \citep{chau2014} operating on scales of seconds and hundreds of meters.

The renewed focus on understanding these types of echoes has underscored the broader need to better measure several key meteor trail  parameters required in common to model both transverse and radial scattering. In particular, the initial trail radius and functional form of the  electron distribution are central to understanding the time series amplitude development of radar signals reflected from meteor trails \citep{Baggaley2009}.

Historically, transverse scattering has been the mode of choice for radar meteor detection, but during the past two decades radial scattering from head echoes has become more common \citep{Close_et_al_head_echo_scattering} as it offers better spatial and temporal resolution. Transverse scattering, however, remains a much more sensitive technique than radial scattering from a meteor head echo \citep{Baggaley2009}, producing cross sections up to six orders of magnitude higher for the same meteor as reflection occurs from a much larger segment of the trail. For example, by making use of this power advantage, \citet{SCHULT_2020_Specular_meteor_micro_meteor_limit} performed a specular observation campaign using a high power large aperture radar by  pointing the beam 30$^\circ$ off zenith; they were able to observe meteors as faint as 16$^{mag}$, some four to five astronomical magnitudes fainter than the head echo limit for the same system \citep{SCHULT_2017_head_echo_mass,Brown2017}. The measurements revealed a mean velocity peak for the sporadic meteors at around 17-18 km/s and were able to detect meteors approaching the ablation limit ($10^-6$g at 12 km/s), providing some of the only radar flux estimates at these sizes.

For the measurement of meteoroid flux using transverse scattering, the initial radius of the ionization column is a critical parameter. It can lead to destructive interference for echoes when the trail radius approaches the radar scattering wavelength \citep{Greenhow1960, Campbell-Brown2003}. This in turn places strong bounds on the detectability of backscatter echoes with height and is a major bias in meteoroid flux estimation \citep[e.g.,][]{Ceplecha1998}. However, work to date shows large scatter in the apparent initial radius between individual meteors at fixed heights \citep{Jones2005a}, making absolute radar flux measurements challenging and often discordant with other techniques  \citep[e.g.,][]{Greenhow1960}. Moreover, the effective initial radius of a meteor trail is an important factor in understanding the physics of meteoroid ablation and subsequent diffusion \citep{Jones:1995_theory_initial_trail_radius} in addition to being a key parameter in detailed numerical models of meteor plasmas \citep{MARSHALL_2017_CMOR, Dimant2017a, Dimant2017b}.

In atmospheric studies, the decay times of underdense meteor echoes from backscatter measurements are a widely used technique for inferring ambipolar diffusion coefficients and, ultimately, temperature in the mesosphere \citep{HOCKING:2001_SKiYMET,Stober:2017_winds_temps,Laskar_2019_anomalous_diffusion,Kero2019}. But echo decay times in the lowest part of the mesosphere show an anomalous apparent increase in the diffusion coefficient  whose origin remains unclear. One suggestion \citep{Younger2014} is that this relates to deionization and hence could be probed through modeling the time series of received echo power. Variations of the electron distribution from the usually assumed Gaussian profile may also result in errors in the diffusion coefficient \citep{Jones1990}.  Although several early works considered non-Gaussian electron distributions \citep[e.g.,][]{Eshleman_1955_theory}, no experimental studies examining possible electron distributions other than Gaussian have been performed for transverse scattering echoes. However, such a sensitivity study has recently been conducted for head echoes by \citet{MARSHALL_2017_CMOR}.

In more general terms, the conversion from received scattered signal at the radio receiver to electron line density (and ultimately meteoroid mass) depends on detailed calculation of the electromagnetic reflection coefficient from the trail \citep{Kaiser_and_Closs_1952_theoryI, POULTER_and_Baggaley_1977_full_wave_scat}. This depends on the assumed functional form of the electron distribution within the trail. While this has historically been taken to be a Gaussian \citep{McKinely:1961}, recent modeling work \citep{Sugar:2018_plasma_model_theory_and_simulation} suggests that the distribution may be substantially non-Gaussian, echoing earlier results \citep{Jones:1995_theory_initial_trail_radius}.

A unifying theme for measurements of the initial radius and electron distribution in transverse scattering studies of meteor echoes is that they yield wavelength-dependent effects. The initial radii of trails, in part, determines the underdense echo power as a function of wavelength \citep{Campbell-Brown2003}, with larger radii and shorter wavelengths producing lower echo amplitudes. Similarly, the time constant for amplitude decay is inversely proportional to wavelength squared \citep{Jones:1995_theory_initial_trail_radius}, independent of the initial electron distribution. Assuming the trail evolution is dominated by ambipolar diffusion, a single diffusion coefficient should be measurable at multiple frequencies for any one trail. More generally, the instantaneous radial distribution of electrons in the cylindrical trail determines the reflection coefficient at each wavelength and hence the observed amplitude. As a result, the amplitude-time profile of common echoes simultaneously recorded at multiple wavelengths provides a potentially powerful means of probing the initial radius and total electron trail content in addition to constraining the possible functional forms for the electron distribution.

Motivated by the work of \citet{Jones:1995_theory_initial_trail_radius}, who simulated the initial electron distribution in the trail and found noticeable deviations from a pure Gaussian distribution, here we numerically compute reflection coefficients for transverse meteor echo scattering for a number of electron distributions as a function of trail radius and compare with measurements. Our goal is both to experimentally establish the most probable radial electron distribution and to estimate the effective initial trail radius for each event. For the first time we have been able to make simultaneous measurements of echoes at three frequencies where the complete trail geometry in each case is known. We are therefore able to compute on a pulse-to-pulse basis the model amplitude-time evolution using the known height, speed, polarization angle, and range. With these data it is possible to estimate what fraction of echoes show received amplitude-time profiles that agree with established theory (excluding chemistry) for a range of electron distributions, initial radii, and diffusion coefficients.

\section{Previous work}
Early studies leveraged the wavelength dependence of transverse scattering by statistically comparing backscatter meteor echo measurements at different frequencies. A large disparity between predicted height distributions of faint meteor echoes and those measured by radar was first noted by \citet{Greenhow1960}, the discrepancy being larger at higher frequencies. One explanation proposed was that the initial trail radii were comparable to the wavelengths being used \citep{Manning1958} resulting in large echo amplitude attenuation from echoes at higher altitudes. Measurements to address the role of initial radius focused mainly on estimating the size of the initial radius from backscattered echo power, assuming trails had a Gaussian radial electron distribution.

The earliest of these experiments was by \citet{Greenhow1960}, who made simultaneous amplitude-time measurements of common echoes at 17 and 38 MHz. The amplitude ratios were used to estimate initial radii as a function of height, which in turn was estimated based on decay times and a model dependence of ambipolar diffusion coefficient (D) with height. Echo numbers and decay time distributions at these two frequencies were also compared to independent measurements at 70 MHz using theory to estimate the initial radius. They found the initial radius scaled much slower than with the mean free path in the atmosphere and also that the initial radii were of order 1~m at 90~km height, increasing to 3~m at 115~km, much higher than predicted by theory \citep{Manning1958}. They also found no dependence on velocity, a surprising result as ions released during ablation at higher speeds should have lower collisional cross sections \citep{Manning1958} and hence yield larger trail radii.

In contrast, \citet{Kashcheyev1963} found a strong dependence of initial radius with velocity when comparing amplitudes of common echoes observed at 75 and 38 MHz. They estimated initial radii of 0.8~m at heights near 90 km. They note in particular that "The most complex consideration in measurements of this type is the determination and selection of meteor trails" for which theory can be appropriately applied. While this comment was aimed primarily to sorting underdense from overdense echoes, other selection effects are important as well. These include using only echoes showing fresnel amplitude oscillations (removing heavily fragmenting meteors) as well as excluding those that show wind shear or turbulence and therefore produce multiple specular points. This emphasizes the fact that existing theory may only apply to a subset of comparatively well-behaved echoes and that no easy criteria exists to identify these "good" echoes a priori.

\citet{Baggaley_1970_initial_radius} estimated both the height and velocity dependence of initial radius by performing simultaneous echo measurements at 10 and 28 MHz. They selected echoes according to their amplitude-time behavior into underdense, overdense, and transition-type echoes while removing echoes distorted by winds. Heights were estimated from decay times (as was commonly done with almost all early meteor backscatter experiments) and velocity approximately estimated using the amplitude-rise time method \citep{Baggaley1997}. They found that initial radius, $r$, varied with the atmospheric neutral mass density ($\rho$) as $r\propto \rho^{-0.45}$ and by velocity as $r\propto V^{-0.57}$. Most experiments conducted between 1960-1980 found similar dependences, in particular with exponents of less than unity \citep{Jones2005a, Baggaley_1970_initial_radius, Ceplecha1998} a finding widely interpreted as bring due to fragmentation \citep{Hawkes1978}.

This led \citet{Campbell-Brown2003} to explicitly model the effects of fragmentation on the estimation of initial radius. Using a model consisting of discrete, Gaussianly distributed fragment "trainlets" they were able to qualitatively reproduce the trend in measured amplitude ratios of Geminid echoes at two frequencies. They also could reproduce the observed height distribution of Geminid echoes at three frequencies. However, significant scatter in relative amplitude values even at fixed heights remained. The solutions derived were not unique, suggestive of processes more complex than the simple fixed fragmentation model used.

A continuation of this work by \citet{Jones2005a} used dual frequency measurements of simultaneous sporadic echoes by the Canadian Meteor Orbit Radar (CMOR) \citep{Jones2005, Brown2008}. They found a very weak speed dependence on initial radius and a significant scatter in the amplitude ratios across all speeds for fixed heights (and vice versa).

The current study is an extension of the earlier works of \citet{Campbell-Brown2003} and \citet{Jones2005a} and also uses the CMOR system, but with several significant differences. First, those original studies were confined to using measurements at only 29 and 38 MHz, despite data also being gathered on 17 MHz. The reason for this was massive interference at the time by terrestrial maritime broadcasts at 17 MHz. In the intervening years, this band has substantially "opened up" as marine broadcasts have greatly diminished, and it is now possible to regularly measure simultaneous echoes at all three frequencies. Secondly, several additional outlying stations have been added to CMOR since the original measurements in 2002 \citep{BROWN:2010_CMOR_shower_catalogue}, making velocity and radiant determination much more accurate. As well, the CMOR detection and analysis pipeline has been optimized for meteor astronomy measurements, rather than atmospheric measurements alone, as was the case in 2002, providing much larger numbers of usable echoes. Finally, custom software has since been produced and refined that allows detailed manual examination of each echo, permitting high quality data sets to be constructed.

\section{Theory - Scattering from cylindrical plasma distributions}
Meteors entering the Earth's atmosphere with a sufficiently high kinetic energy form a plasma column. These plasma trails are easily detected by transverse scatter radars. Here we build from the full wave scattering theory of  \cite{POULTER_and_Baggaley_1977_full_wave_scat} with some revisions. This so-called full wave scattering theory is a first principle approach for deriving reflection coefficients for scattering a linearly polarized electromagnetic wave from an infinitely long plasma cylinder.\\

In this theory, meteor trails are formed instantaneously as an ambipolar diffusing plasma with initial radius $r_0$. The temporal evolution is governed solely by ambipolar diffusion, and, hence, the plasma distributions should follow the radial diffusion (cylindrical coordinates) equation:
\begin{equation}
\frac{\partial n}{\partial t}=\frac{D}{r}\frac{\partial n}{\partial r}\Bigg(r\frac{\partial n}{\partial r}\Bigg)~~,\label{Diffusion_equation}
\end{equation}
where $n$ is the electron volume density ($e^-/m^3$), $D$ is the ambipolar diffusion constant, $r$ describes the distance from the trail axis at time $t$. Assuming a Gaussian electron distribution the radius $a$ of the trail at time $t$ temporally evolves as described by \citep{Herlofson_1951,Kaiser_and_Closs_1952_theoryI,Eshleman_1955_theory};
\begin{equation}
a^2=r_0^2+4Dt~~~.
\end{equation}
\cite{Eshleman_1955_theory} was among the first to present a more generalized scattering theory for transverse meteor geometries including multidimensional shapes and plasma distributions. Recently, several studies investigated different radial plasma distributions to determine their influence on the scattering from meteor head echoes \citep{Marshall_and_Close_2015_FTDT,MARSHALL_2017_CMOR} using a Finite Difference Time-Domain (FDTD) model. Previous studies from \cite{Close_et_al_multi_freq,Close_et_al_head_echo_scattering} already invoked different plasma distributions to estimate the reflection coefficients for meteor head echoes.  In this work we compute reflection coefficients for transverse scattering using three different plasma distributions inspired by the recent work of \citet{MARSHALL_2017_CMOR} related to meteor head echoes. These functions include a Gaussian, exponential-parabolic, and 1-by-r$^2$ radial electron distribution. \\

The Gaussian electron distribution takes the following form:
\begin{equation}
n(r,t)=\frac{q}{\pi a^2}\cdot e^\frac{-r2}{a^2}~~.
\end{equation}
The exponential-parabolic distribution is given by:
\begin{equation}
n(r,t)=\frac{2q}{\pi a}\frac{2 e^{\frac{\pi r}{a}}}{e^{\frac{2\pi r}{a}}+1}~~.
\end{equation}

The 1-by-r$^2$ distribution from \citep{MARSHALL_2017_CMOR} is defined by:
\begin{equation}
n(r,t)=\frac{2q}{\pi^2 a^2} {\frac{1}{1+r^2/a^2}}~~.
\end{equation}

We also explored the recently proposed lateral Sugar model (LSM) or Dimant Oppenheim model (DO) \citep{Dimant:2017,Sugar:2018_plasma_model_theory_and_simulation} where the plasma distribution is approximated by the following equation (we refer to this model as 1-by-$r^3$-plasma distribution):
\begin{equation}
n(r,t)=\frac{q 3^{\frac{3}{2}}}{\pi^2 2 a^3} \frac{1}{1+r^3/a^3}~~.
\end{equation}
All plasma distributions are normalized to the value of the integral taken from zero to $\infty$ over the electron volume density; these results are then translated into an equivalent electron line density $q$, which is what is directly computed from echo amplitudes for transverse scatter radars.\\

Assuming a plane incident wave and omitting the time varying factor ($e^{-i\omega t}$) the dielectric function can be written as:
\begin{equation}
\kappa(r)=1-\frac{n e^2}{\epsilon_0 m \omega^2 (1+i \nu / \omega)}~~~.\label{dielectric_func}
\end{equation}
Here $\epsilon_0$ is the vacuum or free space permittivity, $\omega$ is the incident wave angular frequency, $\nu$ describes the total electron angular collision frequency, $m$ is the electron mass and $e$ denotes the unit of elementary charge.\\

The total collision frequency in eq.~\ref{dielectric_func} is an important quantity that can greatly affect the reflection coefficients and requires some additional information and assumptions of the atmospheric state. In this study we apply the collisional model following \cite{Marshall_and_Close_2015_FTDT}. They estimated the electron-neutral collision frequency based on MSIS \citep{Hedin_1991_MSIS} and used the relations given by \cite{Callen_2006} to estimate the electron-electron and electron-ion collision frequencies.\\

A detailed description of the full wave scattering theory is outlined in \cite{POULTER_and_Baggaley_1977_full_wave_scat}. Here we summarize the basic steps of the full wave scattering theory; the interested reader is referred to \cite{POULTER_and_Baggaley_1977_full_wave_scat, POULTER_and_Baggaley_1978} for a full derivation of the equations and some early comparison with measurements.

For cold, collisional and unmagnetized plasma the electric and magnetic field components of a radio wave inside the ionized plasma column of a meteor trail are described by the differential form of Maxwell's equations:
\begin{eqnarray}
\nabla \times \mathbf{H} & = & -i \omega \kappa \epsilon_0 \mathbf{E}\\
\nabla \times \mathbf{E} & = & -i \omega \mu_0 \mathbf{H}~~~,  \nonumber
\end{eqnarray}
where $\mu_0$ is the free space permeability and the current density $J$ from Ampere's-Maxwell law is included in the $\kappa$. Furthermore, a direct consequence of the ambipolar diffusion of the trail is that $\nabla \cdot \mathbf{E} =0$ and $\nabla \cdot \mathbf{H} =0$. However, this assumption is only valid as long as the trail life-time is much shorter than the timescale for electron-ion recombination, which is generally the case for underdense meteors above 85 km altitude \citep{Younger2014, Baggaley1979}. Thus, the fields inside the column are given by:
\begin{eqnarray}\label{Maxwell}
\nabla^2 \mathbf{E} + k^2\kappa \mathbf{E} & = & 0\\
\nabla^2 \mathbf{H} + (\nabla \times \mathbf{H}) \times \frac{\nabla \kappa}{\kappa}+k^2\kappa \mathbf{H} & = & 0~~~,  \nonumber
\end{eqnarray}
where $k=2\pi/\lambda$ and $\lambda$ is the wavelength of incident wave in free space. We then transform eq.~\ref{Maxwell} into cylindrical coordinates from (x,y,z) to (r,$\phi$,z) where the z-axis is aligned with the meteor trail and $r$ measures the distance from the trail at an angle $\phi$. Following \cite{POULTER_and_Baggaley_1977_full_wave_scat}, we expand the electric and magnetic fields inside the column by a Fourier series:
\begin{eqnarray}
\mathbf{E} & = & \sum_m a_m P_m cos(m \phi)\\
\mathbf{H} & = & \sum_m b_m T_m cos(m \phi)~~~.
\end{eqnarray}
Inserting this approach into eq.~\ref{Maxwell} one obtains a series of differential equations for $P_m$ and $T_m$. The two Fourier coefficients $P_m$ and $T_m$ are related to the polarization direction of the incident plane wave, which can be either parallel or transverse to the trail orientation.\\

For the parallel case, the electric field inside the column should follow the second order differential equation:
\begin{eqnarray}\label{parallel_eq}
\frac{d^2 \mathbf{P_m}}{d r^2}+ \frac{1}{r}\frac{d \mathbf{P_m}}{d r} + \bigg( k^2\kappa-\frac{m^2}{r^2}\bigg) \mathbf{P_m}&=&0~~.
\end{eqnarray}
The transverse case is given by the differential equation:
\begin{eqnarray}\label{transverse_eq}
\frac{d^2 \mathbf{T_m}}{d r^2}+ \Bigg[\frac{1}{r}-\frac{1}{\kappa}\frac{d \kappa}{d r}\Bigg]\frac{d \mathbf{T_m}}{d r} + \bigg( k^2\kappa-\frac{m^2}{r^2}\bigg) \mathbf{T_m}&=&0~~.
\end{eqnarray}
The incident plane wave outside the plasma column is given by the cylindrical Bessel function (omitting the time dependence):
\begin{equation}
\mathbf{E_{inc}}=e^{ikx}=\sum_m i^m J_m(kr) \cos(m\phi)~~,
\end{equation}
where $J_m$ is the m-th order cylindrical Bessel function. The reflected wave can be written as
\begin{equation}
\mathbf{E_{ref}}=\sum_m i^m t_m H^{(1)}_m(kr) \cos(m\phi)~~~.
\end{equation}
Here $t_m$ is the m-th order reflection coefficient and $H_m^{(1)}$ is the m-th order Hankel function of first kind.\\

The reflection coefficient is then obtained by solving the second order differential equations by numerical integration out to the boundary matching radius, where the fields inside the column should match the fields outside the plasma.\\

The total reflection coefficient is then given by summation over all relevant orders used in the Fourier expansion, which correspond to the m-th order Bessel and Hankel functions. As a result, we obtain reflection coefficients $g_{||}$ for the parallel case and $g_\perp$ for the transverse case, which are given by:
\begin{eqnarray}
g_{||}&=& \sum_m t_m \cos(m \phi) \\
g_{\perp}&=& \sum_m - t_m \cos(m \phi)~~~.
\end{eqnarray}
Assuming a backscatter geometry we set $\phi=180^\circ$.

\section{Numerical solution of full wave models}
\label{sec:model}
The reflection coefficients obtained in this study are based on numerical solutions of eqs.~\ref{parallel_eq} and \ref{transverse_eq} using a 5th order Runge-Kutta algorithm with adaptive step size control \citep{Cash_Karp_numerical_integration}. Runs were performed on an Intel Nehalem EP 2,67 GHz cluster nodes with 12 cores. The code is parallelized using 6 CPUs for each modeled radar frequency. Later the code was moved to UBELIX (University of Bern Linux Cluster). UBELIX permits the allocation of 20 cores to solve in parallel all orders of the Bessel functions per parameter setting.\\

The adaptive step size control in this approach turned out to be essential to ensure numerical stability of the solution. We used an $\epsilon=10^{-12}$ as the fractional error threshold. In particular, the resonances in the transverse case are numerically challenging and computationally intensive to properly characterize. We used a truncation radius or stopping radius for the boundary matching using eq.~\ref{dielectric_func} of $\kappa=0.99999$ for all three plasma models, which we considered as physically equivalent to free space permittivity.\\

The algorithm was validated by reproducing the results shown in \cite{POULTER_and_Baggaley_1977_full_wave_scat} applying the collisional damping described therein. We conducted several sanity checks and were able to reproduce quantitatively the reflection coefficients and phase behavior for both the transverse and parallel cases as given in their work.

\begin{figure}
   \centering
     \resizebox{\hsize}{!}
   {\includegraphics[width=12cm]{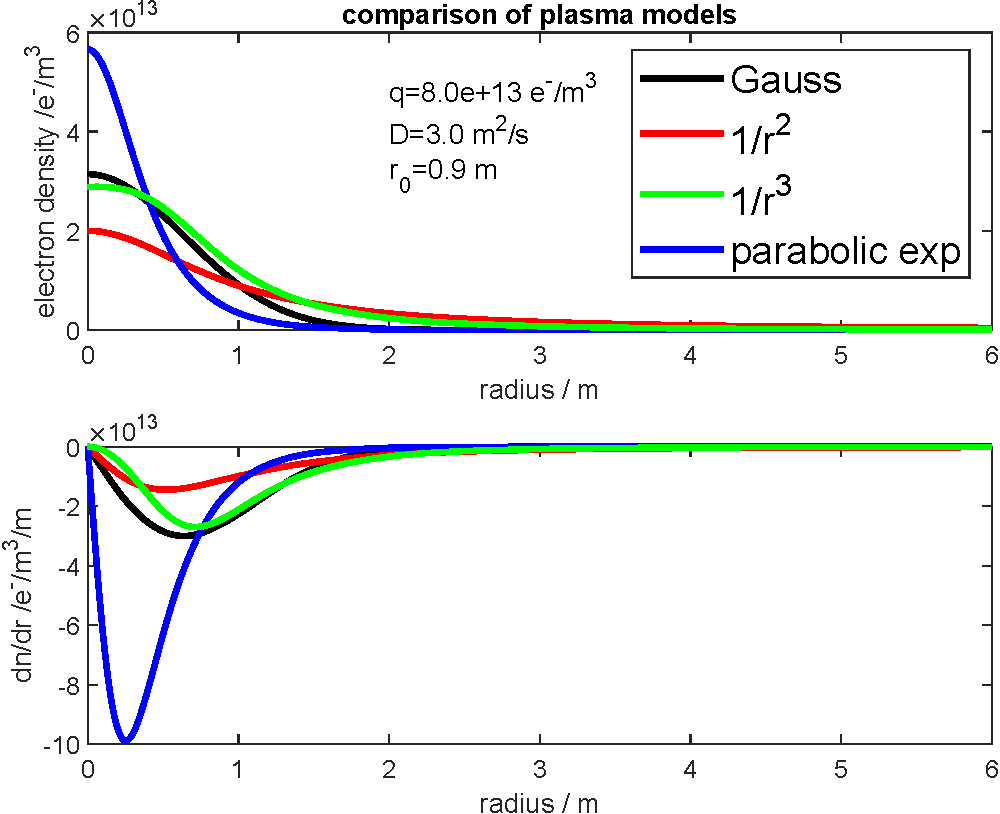}}
      \caption{Radial dependence of plasma distributions. Upper panel: Comparison of the four different plasma distributions and their radial decay with increasing distance from trail axis. Lower panel: Gradient of the dielectric function for each plasma model as a function of radial distance from the trial axis.}
         \label{Fig_plasma_distrib}
\end{figure}

Another important consideration in computing the reflection coefficients for our three different plasma models is the form of the electron density decay with increasing radial distance from the trail axis. In Figure~\ref{Fig_plasma_distrib} we compare this decay for representative values of initial radius, diffusion, and electron line density for the three plasma models. The upper panel shows the electron volume density as a function of distance from the trail axis while the lower panel visualizes the electron density gradient. It is obvious that the parabolic exponential model reaches the highest electron densities and exhibits the largest electron density gradients. The 1-by-r$^2$ model shows a much slower decay with increasing distance from the trail axis and also a much weaker electron density gradient compared to the two other models and can be expected to produce very different scattering characteristics as a result. The Gaussian radial decay model somewhat straddles the other two models, though is closer in numerical behavior to the 1-by-r$^2$ model. It is almost identical to the 1-by-r$^3$ model - we therefore do not explore further fits with 1-by-r$^3$ noting only that they should be virtually identical to the Gaussian model.  The slower radial decay of the electron density of the 1-by-r$^2$ model resulted in it requiring more than an order of magnitude more computational time than the other two models to generate reflection coefficients.\\

For each of the three plasma distributions, we generated look up tables for parallel and transverse reflection coefficients. The computational domain was chosen to cover the expected values for underdense meteors detectable by CMOR and more generally applicable for most VHF frequencies and all-sky antenna arrays. The initial radius ranges from 0.1 cm to 25 m with varying resolutions. From 0.1 cm up to 2 m initial radius, we use a step size of 1 cm, from 2 m to 15 m we apply a step size of 10 cm and for initial radii between 15 and 25 m we use a 1 m step. The electron line densities $q$ range from $q=10^{11}~e^-/m$ to $q=9\cdot 10^{14}~e^-/m$. All intervals are logarithmically scaled for each decade in electron line density. In total we used a domain size of 36 different electron line densities.\\

\begin{figure*}
   \centering
    \resizebox{\textwidth}{!}
   {\includegraphics[width=14cm]{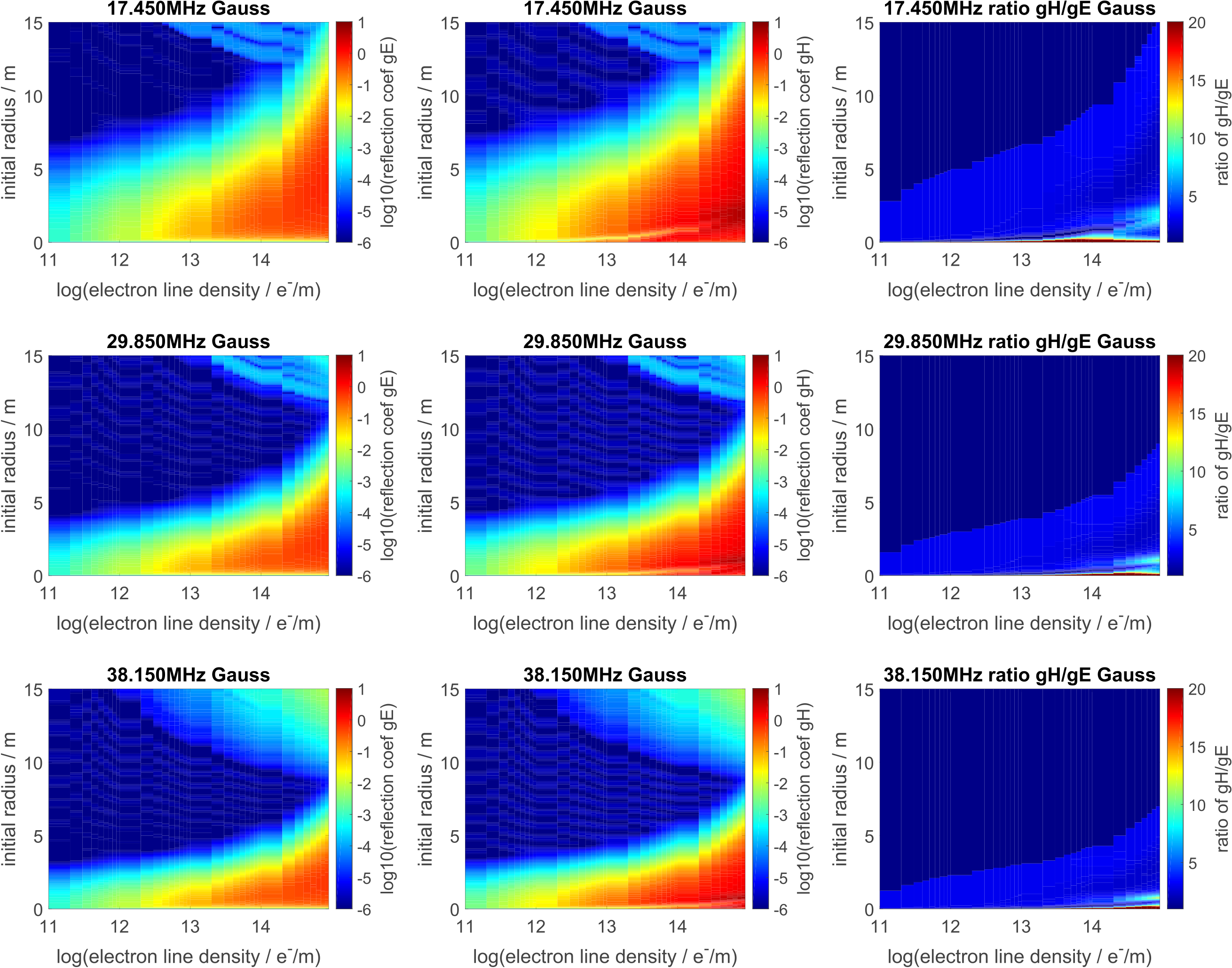}}
      \caption{Parallel (left panels) and perpendicular (central panels) polarization reflection coefficients for a Gaussian plasma distribution for trail radii from 1 cm to 15 m at the CMOR frequencies of 17.45, 29.85, and 38.15 MHz. The x axis shows the electron line density, the y-axis the trail radius half-width and the color contours show the log$_{10}$ of the reflection coefficient. The right column shows the polarization between perpendicular and parallel scattering coefficients.}
         \label{Fig_reflection_coeff_overview_Gauss}
\end{figure*}

Figure~\ref{Fig_reflection_coeff_overview_Gauss} shows our numerical results for parallel and transverse reflection coefficients for the Gaussian radial electron distribution at all three CMOR frequencies. The color contours are in logarithmic scale; effectively only reddish and yellowish areas are visible for typical VHF- all sky meteor radars. The bluish part represents reflections that in practice would be undetectable and corresponds to initial radius sizes much larger than the radar wavelength.  This part also contains a regular pattern analogous to that seen in the optical regime for Mie-scattering reflection coefficients for spherical dielectric particles \citep{hulst1981}.\\

Differences between the parallel and transverse reflection coefficients are most pronounced for larger electron line densities and smaller initial radii. In particular, strong resonance effects are evident in the model at radii where the peak plasma volume density becomes overdense (when $\kappa$=0). Such resonances occur in the transverse case. A more detailed discussion of these effects and the data is given later in the manuscript.\\
\begin{figure*}
   \centering
    \resizebox{\textwidth}{!}
   {\includegraphics[width=14cm]{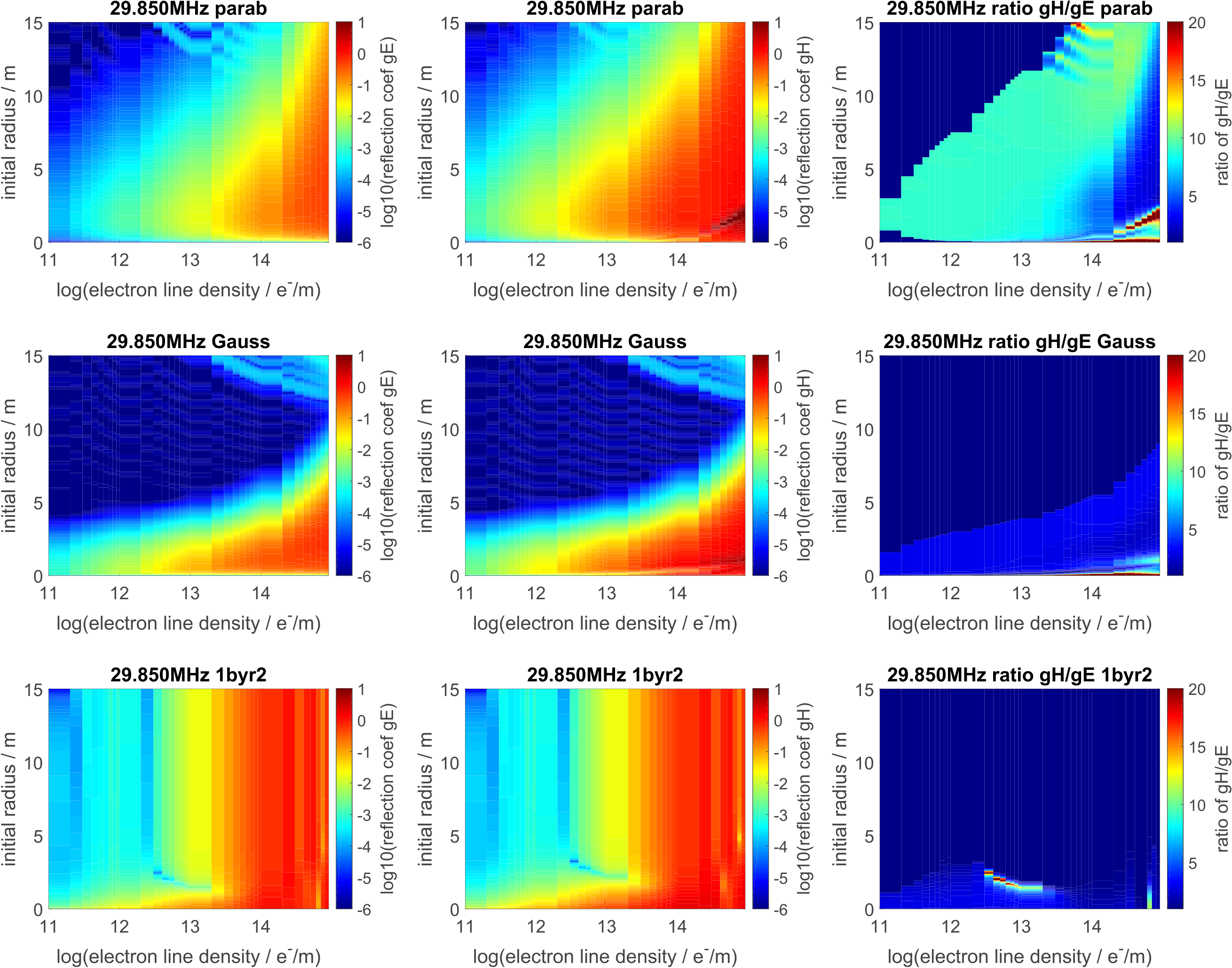}}
      \caption{Parallel (left panels) and perpendicular (central panels) reflection coefficients for parabolic exponential (top row), Gaussian (middle row) and  1-by-r$^2$ (lower row) plasma distributions having initial radii from 1 cm to 15 m at a frequency of 29.85 MHz. The axis and contours are the same as Figure \ref{Fig_reflection_coeff_overview_Gauss}. The right column shows the polarization between perpendicular and parallel scattering coefficients.}
         \label{Fig_reflection_coeff_overview_models}
\end{figure*}

Finally, in Figure~\ref{Fig_reflection_coeff_overview_models} we compare the reflection coefficients for the different plasma distributions at the middle CMOR frequency of 29.85 MHz. The upper panels show the parabolic exponential model, the central two panels show the Gaussian model and the lower panel the 1-by-r$^2$ radial electron distribution. The most obvious difference between the models is the width of the reddish and yellowish areas in the parameter space for small initial radii. The parabolic exponential plasma distribution shows a much slower decay of the reflection coefficient with increasing initial radii compared to the 1-by-r$^2$ model. In addition, the parabolic peak plasma density is more likely to show overdense scattering for small initial radii.  Another notable difference between the plasma distributions is that the parabolic exponential model shows a much more pronounced difference between the parallel and the transverse scattering cases. As a consequence it also shows much stronger resonance effects compared to the other two plasma distributions. The dramatic difference in reflection coefficients for the same meteor trail parameters between the parabolic and two other distributions, in particular, suggests that observational comparisons will be able to clearly distinguish between the parabolic and other radial electron profiles.

\section{Resonance effects for transition echoes}
Resonance effects in transverse scattering have been predicted in many early works, such as \cite{Herlofson_1951,Kaiser_and_Closs_1952_theoryI,POULTER_and_Baggaley_1977_full_wave_scat} and are a direct consequence of the $1\over{\kappa}$ factor in the denominator of the second term of eq.~\ref{transverse_eq}. The real part of the complex permittivity $\kappa$ becomes zero at the transition from the overdense plasma region to the underdense plasma region and, thus, generates a singularity. This singularity is numerically solved by the complex nature of the permittivity $\kappa$. Physically the imaginary part of the complex permittivity $\kappa$ describes a coupling with the ambient atmosphere through electron-electron, electron-neutral, and electron-ion collisions (see \citet{Callen_2006,Marshall_and_Close_2015_FTDT}).\\

From the equations,  resonance effects only occur in the transverse case for a linearly polarized incident wave. Due to the (assumed) infinite extension of the plasma column in the parallel direction, the electrons cannot physically enter into a resonant oscillation according to eq.~\ref{parallel_eq}, whereas in the transverse case the finite characteristic width of the plasma column is on the order of a few meters, and, hence, the electrons in the meteor trail can exhibit a resonant scattering induced by the exciting wave.\\
\begin{figure}
   \centering
     \resizebox{\hsize}{!}
   {\includegraphics[width=12cm]{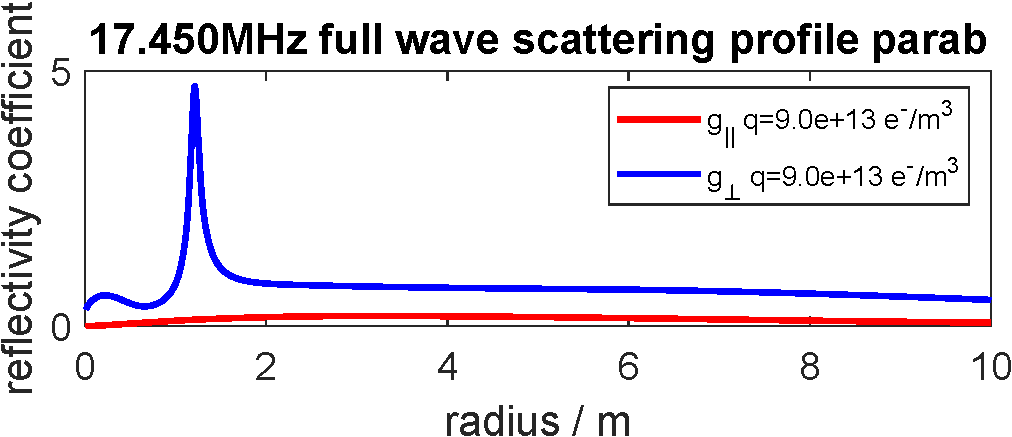}}
     \resizebox{\hsize}{!}
   {\includegraphics[width=12cm]{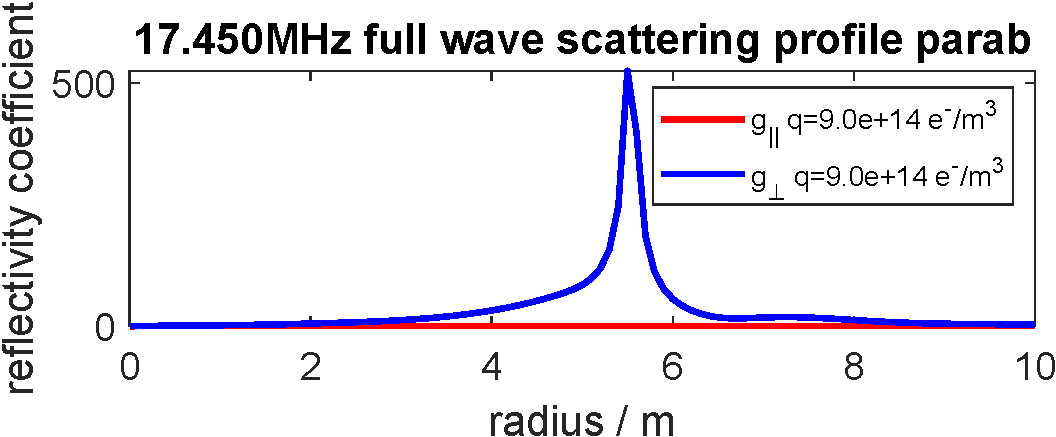}}
      \caption{Reflection coefficients at 17.45 MHz as a function of trail radius showing resonance signatures for a parabolic plasma density, which are extreme cases not found in the observations. The upper panel shows a meteor close to the overdense regime with an electron line density of  $q=9\cdot10^{13}e^-/m$. The lower panel shows an overdense echo with an electron line density of  $q=9\cdot10^{14}e^-/m$.}
         \label{Fig_resonance_parab}
\end{figure}
\begin{figure}
   \centering
    \resizebox{\hsize}{!}
   {\includegraphics[width=12cm]{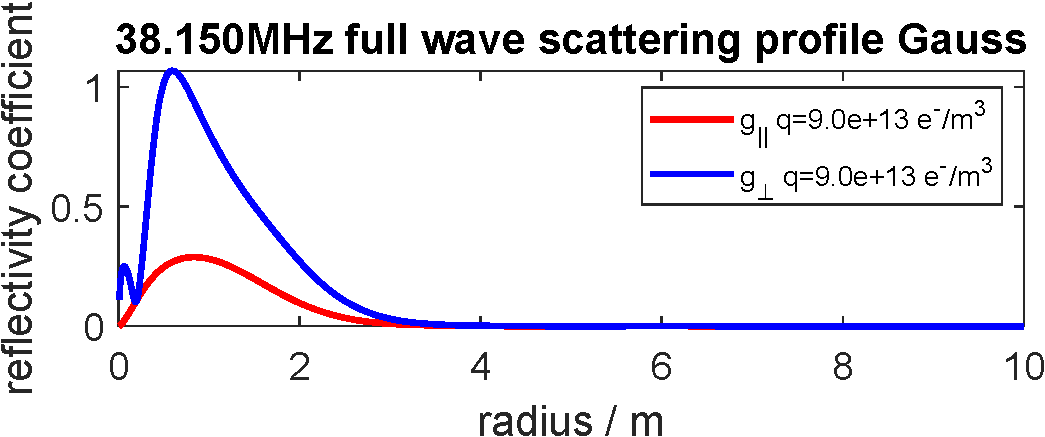}}
    \resizebox{\hsize}{!}
   {\includegraphics[width=12cm]{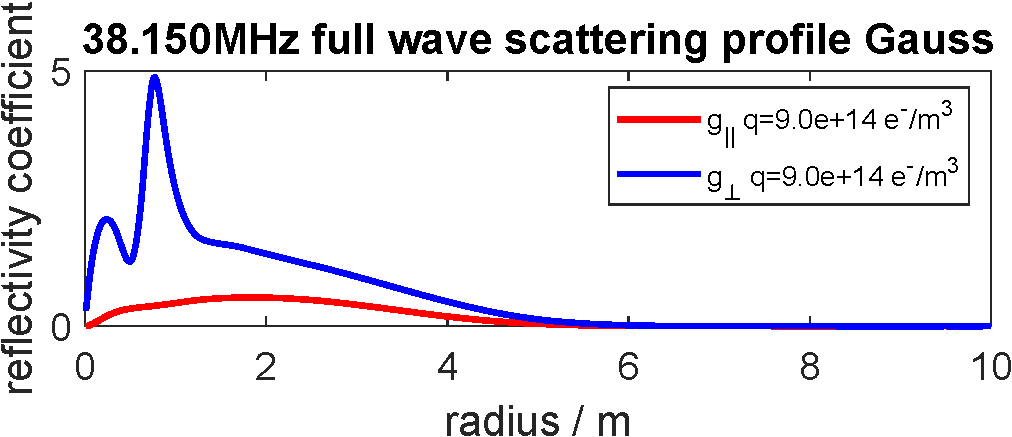}}
      \caption{Reflection coefficient at 38.15 MHz as a function of the trail radius showing resonance signatures for a Gaussian plasma density. The upper panel shows a meteor close to the overdense regime with an electron line density of  $q=9\cdot10^{13}e^-/m$. The lower panel shows reflection coefficient values for a strongly overdense echo having an electron line density of  $q=9\cdot10^{14}e^-/m$.}
         \label{Fig_resonance_gauss}
\end{figure}

The magnitude of the resonance effects depend on the frequency and radial electron distribution, which ultimately defines the transition between an overdense and the underdense scattering regime for a specific frequency. In Fig.~\ref{Fig_resonance_parab} and \ref{Fig_resonance_gauss} we show examples of transition echoes and overdense echoes that show extreme resonance effects. The parabolic exponential plasma distribution in particular predicts extreme resonances at 17 MHz. However, as discussed later we were not able to find examples that could be fit with this plasma model and have found no observational evidence for such large resonances in actual echoes observed at this frequency. Thus, these resonance curves are more of theoretical interest than a practical reflection of the physical conditions in any trail.  The temporal evolution of the signal is produced  as the radial extent of the trail changes with time. Denser plasma distributions, such as the parabolic exponential, tend to evolve significant resonance amplification of the signal, whereas the Gaussian plasma distribution shows a much weaker resonance, more physically consistent with observations for the same electron line densities. However, even for the Gaussian distribution, the larger reflection coefficient for the transverse scattering is still significantly higher than for the parallel case.\\

These theoretical results imply that we are best able to constrain individual trail properties and the radial electron distribution by focussing our experimental comparisons on underdense echoes, which show no clear resonance effects, but do show a characteristic morphology for underdense echoes. The resonance effects mean that the reflection coefficient changes rapidly with increasing electron line density and trail radius in the overdense regime making overdense echoes less amenable to inversions of these quantities (see detailed discussion in \citep{WerykBrown2012}. Furthermore, these results suggest that there is no clear threshold for a certain electron line density to define the transition between a morphologically distinct underdense or overdense echo as has been noted before \citep{Ceplecha1998, WerykBrown2012} as this depends on the angle between the incident wave and the meteor trajectory.

\section{Angular dependence of reflection coefficients on meteor trajectory}
One of the fundamental results of the full wave scattering model is that there is expected to be an angular dependence of the echo morphology on the scattering angle $\delta$; this is the angle between the incident radar plane wave and the meteor trail direction. From the numerical output of the full wave scattering model, we obtain a set of reflection coefficients $g_{\perp}$ and $g_{||}$, which describe transverse and parallel scattering, respectively. The total reflection coefficient for arbitrary scattering angles $\delta$ is then given by the linear combination \citep{POULTER_and_Baggaley_1977_full_wave_scat}:
\begin{equation}
    g=g_{||}\cdot cos^2(\delta)+g_{\perp}\cdot\sin^2(\delta).
\end{equation}
The solution can be generalized to circular polarization. Crossed dipole antennas usually consist of two linear dipoles physically rotated by 90$^\circ$ and fed with a separate signal adding a 90$^\circ$ phase shift between the two antennas. Thus, a circular polarization is described by the superposition of two linear plane waves.

\begin{figure}
   \centering
   \includegraphics[width=8cm]{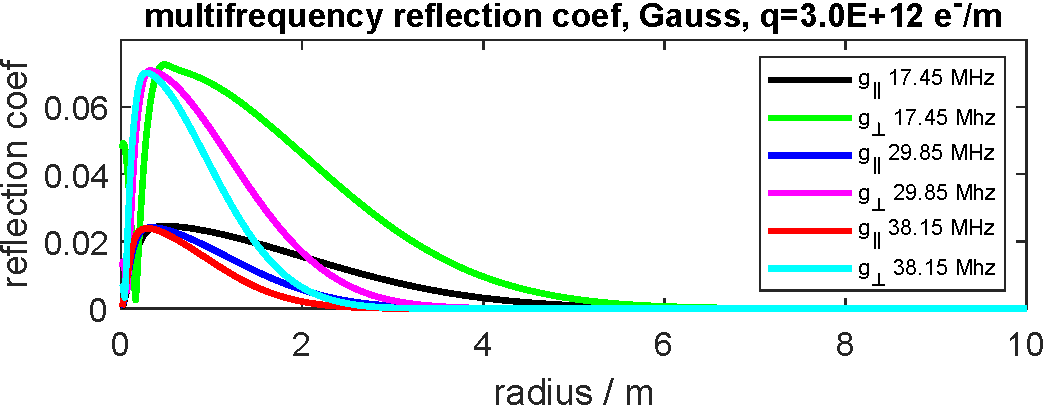}
   \includegraphics[width=8cm]{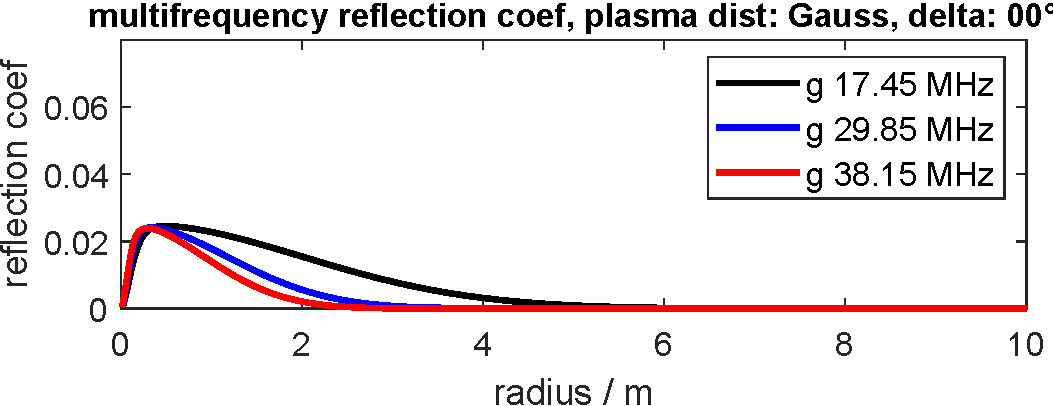}
   \includegraphics[width=8cm]{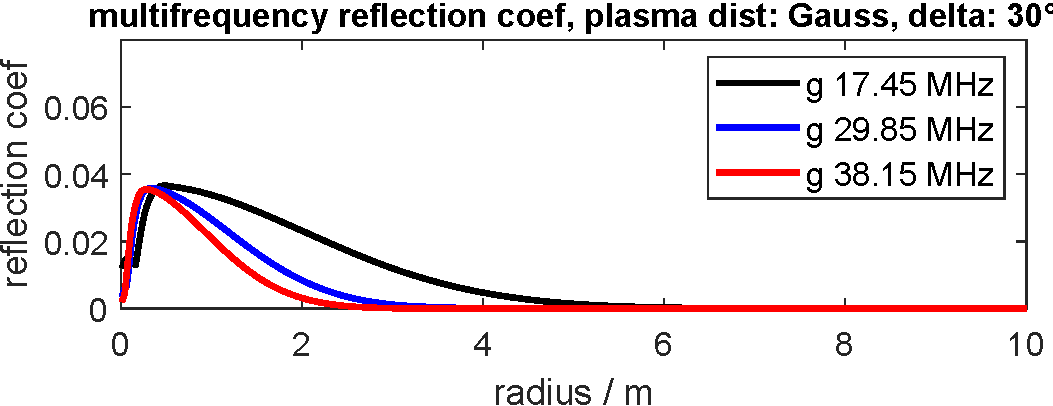}
   \includegraphics[width=8cm]{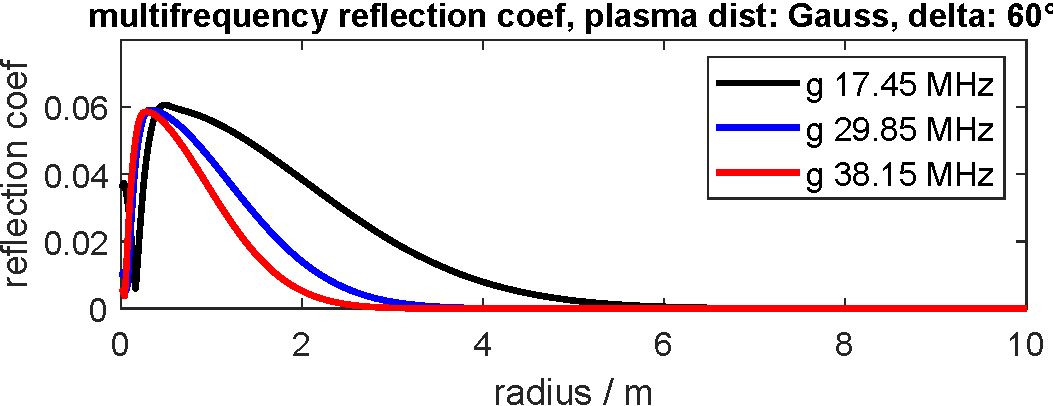}
   \includegraphics[width=8cm]{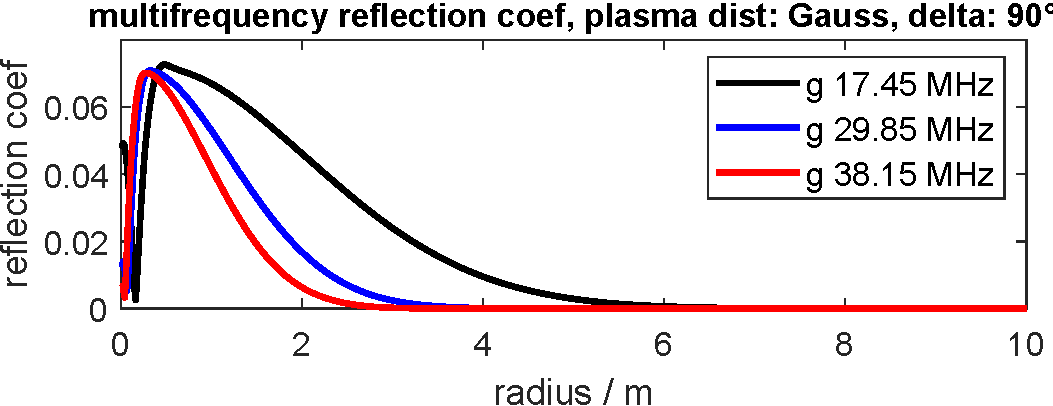}
      \caption{Model estimates of the full wave scatter reflection coefficients for an electron line density $q=3\cdot 10^{12} e^-/m$, using a radial Gaussian plasma distribution at all three CMOR radar frequencies as a function of the trail radius. The upper panel shows the reflection coefficients for the transverse and parallel cases color coded by frequency. The four panels below show the change in total reflection coefficient with initial trail radius (which can be translated into time using a diffusion coefficient) for increasing scattering angles $\delta=0^\circ, 30^\circ, 60^\circ, 90^\circ$.}
         \label{angular_dependence}
\end{figure}

Knowing the scattering angle $\delta$, it is then straight forward to predict the reflection coefficient for the same meteor using a fixed ambipolar diffusion and electron line density as a function of trail radius. In Fig.~\ref{angular_dependence} we show five panels demonstrating the change in the reflection coefficient with scattering angle at the three CMOR frequencies. The total reflection coefficient, $g$, is shown in the lower four panels at differing scattering angles. This shows that a significant difference is expected in both the maximum intensity and rate of signal fading for the expanding trail. The plots also demonstrate that while the 29.85 MHz and 38.45 MHz reflection coefficient profiles look similar, at 17.45 MHz there are much larger differences in intensity and duration. This underscores the frequency dependence expected in meteor signal morphology, a result well documented in observations \citep{Steel1991a}.

\section{Observations and data collection}
\subsection{Canadian Meteor Orbit Radar hardware and experimental configuration}
Data for this study were collected using CMOR, a triple-frequency radar that simultaneously collects multi-station echoes such that full velocity vectors are computed for a subset (of order 4000-5000 echoes) each day. CMOR was chosen for this study as it combines simultaneous measurement of multifrequency echo power with trajectory information providing the main parameters needed to fit a full-wave scattering model to individual events.

Details of the hardware, antenna patterns and basic experimental setup for CMOR have been described in detail by \citet{Webster:2004_CMOR_I}, \cite{ Jones2005}, \citet{ Brown2008}, and \cite{BROWN:2010_CMOR_shower_catalogue}. The detection, analysis and trajectory algorithms and associated software pipeline details are summarized in \citet{Jones2005}, \citet{WerykBrown2012} and \citet{Mazur2020}. Here we provide only a brief summary of the main features of CMOR pertinent to the current work; we refer the interested reader to the foregoing references for greater detail.

CMOR is located near Tavistock, Ontario, Canada (43.264$^\circ$ N, 80.772$^\circ$ W). The CMOR system consists of three synchronized meteor radars operating at 17.45, 29.85 and 38.15 MHz. The 17 and 38 MHz transmitters have a peak power of 6 kW while the 29 MHz system has a peak power of 15 kW. Each radar system uses a triple element, vertically directed Yagi linearly polarized transmitting antenna with directivity of 7.6 dBi producing a beam with a half width half maximum to the half power points of 30$^\circ$. All elements are aligned to a local azimuth of 343 degrees (measured clockwise relative to geographic north), hence all transmit antennas have the same relative gain and polarization per echo. The transmitters produce Gaussian tapered pulses of 75 $\mu$sec duration, corresponding to a round-trip pulse length of 12 km, and use a 532 Hz pulse repetition frequency, permitting unambiguous range measurement from 18 - 255 km.

Each frequency has five, two-element vertically directed linear yagi antennas arranged in a Jones 5-antenna cross array for reception. In this layout, the receive antennas are separated by 2 $\lambda$ and 2.5 $\lambda$ spacing \citep{Jones:1998_interferometer} with the crossed array axis also aligned along  local azimuth of 343 and 73 degrees relative to geographic north. The two element receive antennas have a directivity of 6.3 dBi and a beam with a half width to the half maximum power points of 45$^\circ$. The resulting echo distributions are broadly all-sky, with peak echo density at an elevation of 40$^\circ$ at an apparent azimuth of 73$^\circ$ and 253$^\circ$.

The accuracy of echo interferometry, found through comparison with simultaneous optical measurements, is typically of order 0.7 -  0.8$^\circ$ \citep{WerykBrown2012} with the majority of echoes having uncertainties below 1$^\circ$. The 29.850 MHz system also has 5 passive remote sites located 5 - 20 km from the main site for reception only, enabling measurement of  time-of-flight velocities \citep{Taylor1991, Jones2005, weryk2012, WerykBrown2012,Mazur2020}. Configuration details for the radar experiment are shown in Table \ref{tab:cmor}.

\begin{table*}[t] 
	\caption{Parameter configurations of CMOR for triple frequency experiments.}
	\label{tab:cmor} 
	\centering 
	
	\begin{tabular}{l l} 
	\hline\hline 
	
Quantity  & Description \\ 
	\hline 
	Location & 43.264$^\circ$N, 80.772$^\circ$W, 324m (WGS-84)\\
Frequency (MHz) & 17.45 , 29.85, 38.15 \\
Pulse duration & 75 $\mu$sec\\
Pulse Repetition Frequency & 532 Hz \\
Range sampling interval & 18 - 255 km \\
Peak Transmitter power (kW) & 6, 15, 6 \\
Range accuracy & $<$ 0.3 km \\
Range resolution & 3 km \\
Noise floor (dBm) & -98, -109, -113\\

\hline 
	\end{tabular}
\end{table*}

\subsection{Data collection, selection, and analysis }
Although the systems are primarily used for astronomical observation of the sporadic meteoroid complex \citep{Blaauw:2011_sporadic_meteors} and meteor showers \citep{BROWN:2010_CMOR_shower_catalogue}, the underlying systems are based on the SkiYMET meteor radar and hence the standard analysis software for SKiYMET systems is running on the systems \citep{HOCKING:2001_SKiYMET} and used for atmospheric studies of MLT (Mesosphere/lower Thermosphere) dynamics \citep{McCormack_2017_MLT_winds}. However, for the data analysis in this study we use a custom  data reduction pipeline optimized to detect meteors of astronomical interest, many of which are unsuitable for wind analysis. The details of the detection pipeline can be found in \citet{weryk2012, WerykBrown2012} and \citet{Mazur2020}. Here we briefly review the basics of the detection pipeline of significance to our study and focus on the selection criteria used to associate common triple frequency echoes and the characteristics required to include them in our analysis.

The initial detection of echoes is performed independently using the streamed raw data on each frequency. A possible echo is declared if an incoherent stack (across all five receivers) of 14 pulses exceeds the noise background by 9dB relative to the standard deviation of the noise. As the transmit pulse length is 4 times the range sampling interval, a typical echo is detected in more than one range gate. In these cases only the echo with the largest amplitude is retained; any echoes within two range gates are ignored for the duration of the "central" echo. Echoes with duration over 5.6 sec are ignored; this efficiently removes noise and non-specular echoes at the expense of a small number of larger overdense echoes. Echoes with apparent heights lower than 70 km or higher than 110 km are removed from further analysis.

For each frequency, the full list of echo events are compared in time and range. Potential common events are associated if the echo is within two range gates of another echo at a different frequency and all occur within 0.1 seconds. Thus only echoes contained inside a 6 km range window on all three frequencies that are within 53 pulses of one another in time are associated.

Once three echoes at different frequencies have been range and time associated, a further check is made that a trajectory is available for the triplet echo. This is done by checking that the common echo detected at 29 MHz has a multi-station velocity solution. This information is used to then define the speed, radiant, and apparent path on the plane of the sky. The apparent path is then used to compute the polarization (scattering) angle $\delta$ between the linearly polarized radar wave and the trail orientation. Finally, the apparent interferometric solution from 29 MHz is used to compute the transmit and receiver gain of each system in the echo direction, and, together with receiver absolute power calibrations (see \citep{WerykBrown2013}), an absolute receive power for each echo is computed per pulse. In practice, we find that the number of detected triplet echoes is strongly dependent on polarization. About 25 triplet echoes with polarization parallel (within a 3 degree tolerance) meet all of the foregoing criteria each day. In contrast, on average about 100 daily triplet echoes with polarization within 3 degrees of perpendicular to the trail orientation are detected.

To select our samples we chose the month of July 2018 as it was relatively free of noise interference across all frequencies and all radar systems were operating stable throughout the entire month. Our procedure was to correlate all echoes from the beginning of the month and then manually examine all triplets chronologically. We divided the data into two groups: echoes having polarizations parallel to the trail axis (within 3 degrees) and those with polarization perpendicular to the trail (also within 3 degrees). Our goal was to manually select 50 "good" echoes in each of the two polarization categories.

Manually examined events were rejected if any of the three frequencies were saturated for any portion of the echo. Events were also rejected if any of the amplitude-time profiles showed multiple maxima after the peak echo power location or unusually slow rise times (indicative of a non-specular echo). Triplet echoes meeting all the foregoing were then examined to determine if their time of flight estimated speeds agreed with the speed estimated from the pre-t$_0$ technique \citep{Mazur2020} to better than 10\% and were retained if this condition was met.

Triplet echoes meeting all the foregoing requirements were then examined and a manual fit attempted. Here the full wave scattering model described in Section \ref{sec:model} was implemented as a lookup table and fits to individual echoes interpolated from the tabular values. On a pulse-to-pulse basis a synthetic amplitude-time profile was computed at all three frequencies on a common time-base using the measured polarization angle, speed, height, range, specular time and echo location within the gain beam as well as the actual transmitter power. Modelled echo profile at all three frequencies were then modified by varying the value for the ambipolar diffusion coefficient, (D), the initial radius, r$_0$, and the electron line density (q). By varying all three of these parameters, the modelled amplitude-time was simultaneously fit on all three frequencies. The goal was to find one set of D, r$_0$, q values that matched the complete amplitude profiles (and particularly the peak power) simultaneously at all frequencies.

Initially, this fit was performed for Gaussian, parabolic and 1-by-r$^2$ radial electron distributions. It quickly became apparent that none of the echoes could be fit by the parabolic-exponential function; in several hundred attempted fits we found not a single case where the parabolic-exponential radial distribution produced a satisfactory fit at all three frequencies simultaneously. As a result, we dropped this electron distribution from further consideration and focused on only the 1-by-r$^2$ and Gaussian distributions. We did find that a minority of echoes for which a fit was possible (about 10\%) could be fit by the 1-by-r$^2$ radial distribution, though it was often only slightly better than the Gaussian distribution. At this stage, it was decided to focus on obtaining robust fits using the Gaussian distribution alone as it clearly produced the largest number of good fits.

Our procedure amounts to selecting potentially good echoes using the various criteria already outlined and then assuming that our model of trail evolution having a single value of ambipolar diffusion, initial radius and electron line density fully explained the change in reflected power with time. In reality, other effects are also in play, such as chemistry (which may remove electrons from the trail) and fragmentation, both of which will affect the power profile \citep{Baggaley2002}. In these instances, we do not expect to find a solution using our model that fits all three frequencies. We also dropped meteor signals that did not follow the classical morphology indicating fragmentation, which is often also related to missing Fresnel oscillations, though we did not use the presence or absence of Fresnel oscillations in the amplitude profile as criteria for inclusion. Optical observations have indicated that about 90\% of all meteors show signs of different types of fragmentation down to millimeter-sizes \citep{Subasinghe_2016_fragmentation,VIDA_2021_fragmentation_density}, and since these measurements overlap in size with our measurements, we suspect fragmentation is the main filter in our selection. As a result, our methodology should be viewed as heavily biased toward echoes exhibiting the simplest behavior and not necessarily representative of the population as a whole. In particular, meteoroids that fragment are likely to be strongly biased against selection using our acceptance criteria.

Finding 50 echoes with parallel polarization for which fits were possible required examining 913 triple frequency echoes. To find the same number for transverse polarization required examining 1709 triplets.

\section{Results}
We compared our theoretical derived reflection coefficients as a function of time in development of the trail from the full wave scattering model against actual observations with CMOR as described in the previous section.

To construct the numerical model, we use the known antenna gain pattern of CMOR \citep{WerykBrown2013} and the measured trajectory of the meteors using the multi-static receiver sites \citep{Mazur2020}, to determine the polarization angle of each backscattered echo. Together with the information CMOR provides on the echo height and velocity, an absolute backscatttered power can be computed. Following \cite{Ceplecha1998} the total backscattered power can be expressed by:
\begin{equation}
    P_R  =   \frac{g^2  \lambda^3  G_R G_T  P_T}{32 \pi^4 R_0^3}~~~.
\end{equation}
Here $G_R$ and $G_T$ are the antenna gains on reception and transmission, $P_T$ is the transmitted power in Watts, $R_0$ denotes the range between the radar and the specular point of the meteor trail and $g$ is the reflection coefficient. As specular echoes reflect near-field scattering, we need to also adopt the Fresnel integrals $C$ and $S$ to define the power time profile of a forming echo defined as:
\begin{equation}
    C=\int_{-\infty}^x\cos\Biggr(\frac{\pi x^2}{2}\Biggr)~~ and~~ ~S=\int_{-\infty}^x\sin\Biggr(\frac{\pi x^2}{2}\Biggr)~~,
\end{equation}
where the Fresnel parameter $x$ is given by:
\begin{equation}
     x=\frac{2s}{\sqrt{R_0 \lambda}}~~~.
\end{equation}
The Fresnel parameter $x$ is determined by the wavelength $\lambda$ and the distance from the specular point along the meteor trail $s$ \citep{Baggaley2005}. The meteor velocity is determined by the time of flight approach making use of the multi-static CMOR configuration (see Appendix A in \cite{Mazur2020}) and, thus, the Fresnel parameter $x$ is known for each transmitted pulse. The theoretical echo time-power profile for meteoroids with known velocity is described by:
\begin{equation}
    P_R  =   \frac{g^2  \lambda^3  G_R G_T P_T}{32 \pi^4 R_0^3}\cdot \Biggr( \frac{C^2+S^2}{2} \Biggl)~~~.
\end{equation}\label{theoretical_echo_profile}

The reflection coefficient denotes the fraction of the total electric field impinging on the trail that is reflected back to the receiver; it implicitly incorporates information about $q$, $r_0$ and $D$. Additionally, g will in general vary as a function of time as the trail expands or the electron line density falls.

\begin{figure*}
   \centering
    \resizebox{\textwidth}{!}
   {\includegraphics[width=14cm]{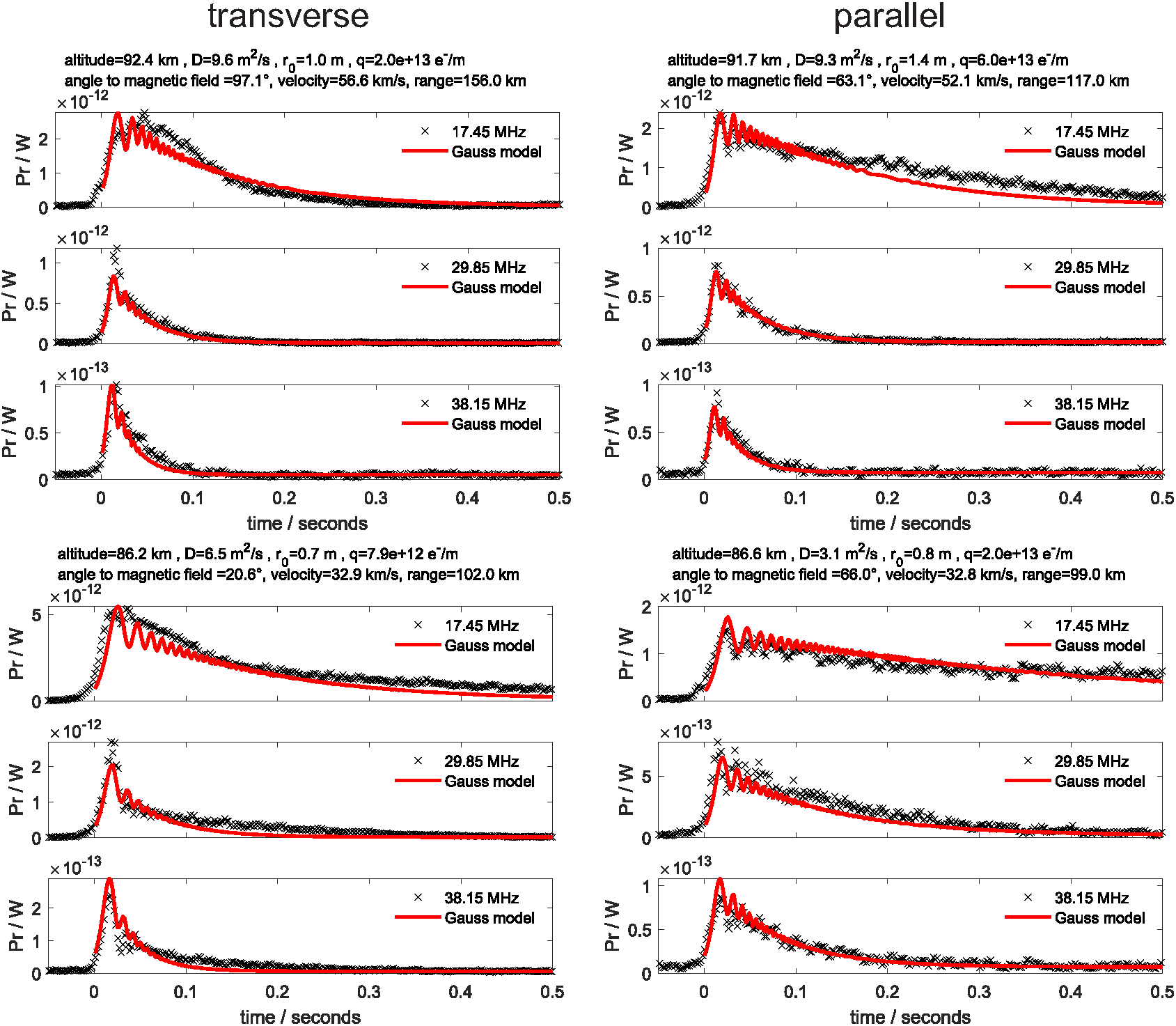}}
      \caption{Examples of echoes for which model fits were possible. The two echoes in the left panels are for transverse polarizations while the two in the right panels are for parallel polarizations. The characteristics of the echoes are given in the title to each plot as are the model fits of D, $r_0$, q. The pulse-to-pulse received power as a function of time are shown as black X's at all three frequencies while the model power as a function of time is shown in red. }
         \label{full_wave_scattering_example}
\end{figure*}

In Figure~\ref{full_wave_scattering_example} we show examples of triple frequency echoes having polarization angles perpendicular and parallel to the trail orientation ie. for transverse and parallel scattering cases. The black crosses are the single pulse echo power at each frequency. The red line describes the theoretical echo profile obtained from eq.~\ref{theoretical_echo_profile}. The optimal choice of the initial trail radius $r_0$, the electron line density $q$ and the diffusion coefficient $D$ were manually selected by varying the parameters to find a best match for all three frequencies using the Gaussian plasma model. As the time between the $t_0$-point and the maximal amplitude is only partially described by the full wave model, we approximate this section by a scaling factor. Strictly speaking, the full wave scattering model is valid only for fully formed trails much longer than the radar wavelength. As a result, the reflection coefficient has half the value at the $t_0$-point compared to the fully evolved trail after reaching the maximum amplitude. \\

The general echo morphology is comparable between the parallel and the transverse cases. However, on closer examination apparent differences in the backscattered power for each frequency are visible. For example, the presence of rapid, large amplitude oscillations due to resonances are visible for many transverse echoes shortly after the specular point. This is not present in parallel scattering echoes. . Another significant difference between the transverse and parallel scattering cases are the decay times, although both examples have comparable diffusion coefficients. This is expected as both types of meteor trails decay in a similar atmospheric environment.\\

In total, we manually identified 50 transverse scattering echoes and 50 parallel scattering events for which model fits on all three frequencies could be found. This represents only 3\% and 5\% respectively of all the triplet echoes manually examined. Although this is just a tiny subset of the CMOR observations, it provides sufficient statistics to evaluate/validate some of the theoretical predictions from the full wave scattering model with actual observations and to examine trends in D and $r_0$. We found only unique solutions for all three parameters incorporating all three frequencies.\\

\begin{figure*}
   \centering
    \resizebox{\textwidth}{!}
   {\includegraphics[width=14cm]{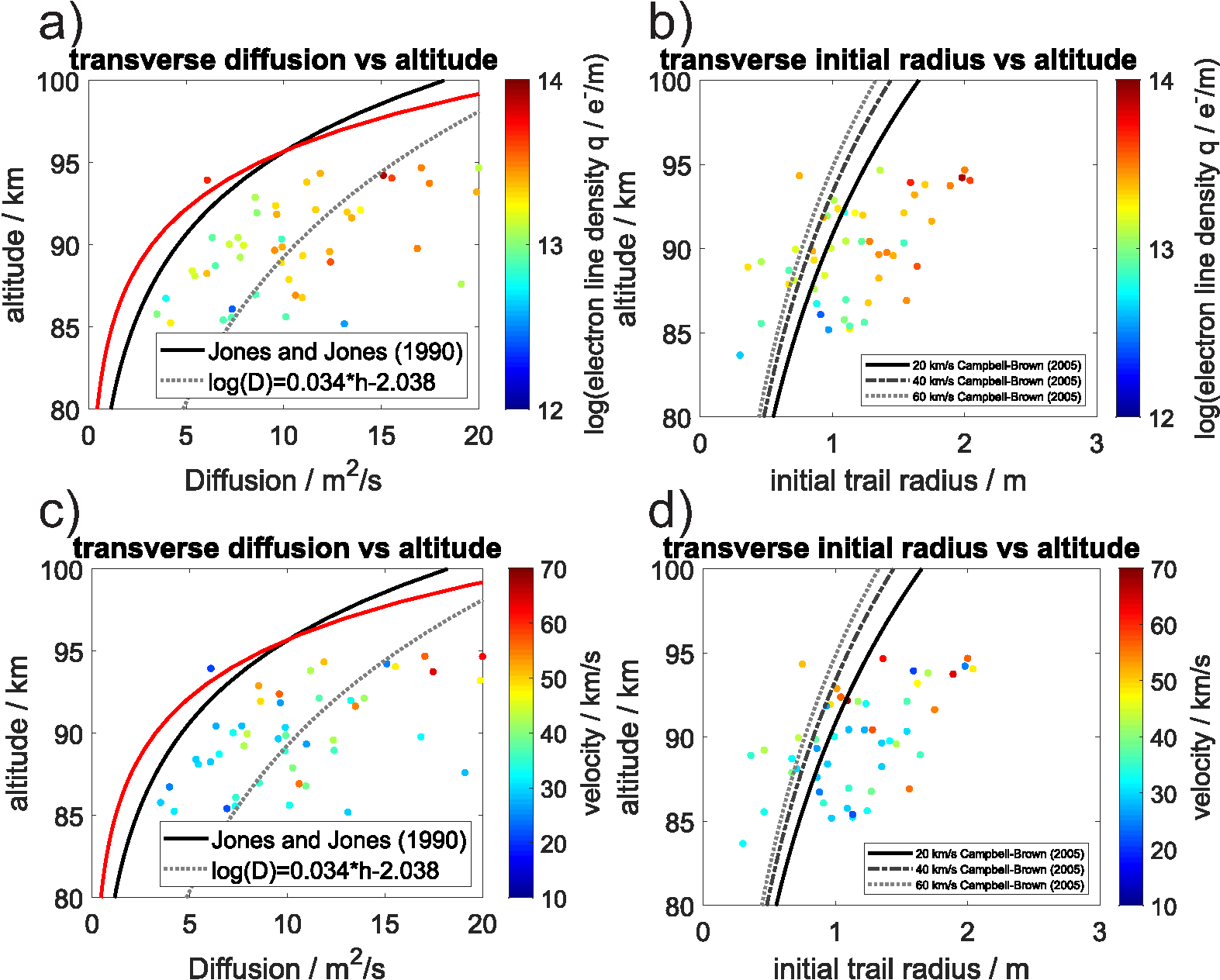}}
      \caption{Comparison of transverse scatter fits. a) Diffusion versa altitude scatter plot with color coded electron line density. b) Initial trail radius versa altitude scatter plot with color coded electron line density. c) Diffusion versa altitude scatter plot with color coded time of flight meteor velocity. d) Initial trail radius versa altitude scatter plot with color coded time of flight velocity. The initial radius as a function of altitude and speed as measured by \citet{Jones2005a} is also shown for comparison.}
         \label{full_wave_scattering_polninety}
\end{figure*}

In Figure~\ref{full_wave_scattering_polninety} we show scatter plots of all our transverse scatter specular echoes displaying a) diffusion versus altitude with color coded electron line density, b) initial trail radius versus altitude with color coded electron line density, while the lower panels c) and d) compare diffusion and initial trail radius with altitude color coded by velocity, respectively.
All four panels show significant scatter indicating only a relatively weak correlation between diffusion and initial trail radius with altitude. There is a clear tendency for fainter meteors (smaller electron line density) to remain detectable at lower altitudes and brighter meteors (larger electron line density) to be more often detected at higher altitudes as expected given the reflection coefficient behaviors as a function of height/initial radius. In addition, the scatter plot in panel a) shows a tendency that fainter meteors correspond to smaller ambipolar diffusion than those closer to the overdense electron line density, again reflecting the internal correlation between line density and initial radius. The diffusion versus altitude scatter plot with color coded meteor velocity also reflects the well-known correlation between ablation height and speed \citep{Vida2018}.

The initial trail radius versus altitude scatter plots in panel b) and d) show a tendency toward larger radii with altitude over the narrow height range of our measurements. While this is roughly consistent with earlier changes in initial radius with height that show a radius doubling every 10-12~km at these altitudes \citep{Jones2005a}, our rate of change with altitude is slightly higher, albeit with significant scatter. Moreover, our absolute initial radius values for transverse scattering events are systematically larger (by about a factor of two) compared to earlier measurements at similar heights \citep{Greenhow1960, Jones2005a}. For transverse scattering, the smallest initial trail radii we observe is 0.5~m, which occurred below 90 km altitude, and the largest is 2~m at an altitude of approximately 95~km. Most of the analyzed meteors had initial trail radii of 1-1.5~m. The trend with height for three speeds as given in the earlier work of \citet{Jones2005a} is shown in the right panels for comparison - our transverse scattering radii are generally higher at each height.\\

\begin{figure*}
   \centering
    \resizebox{\textwidth}{!}
   {\includegraphics[width=14cm]{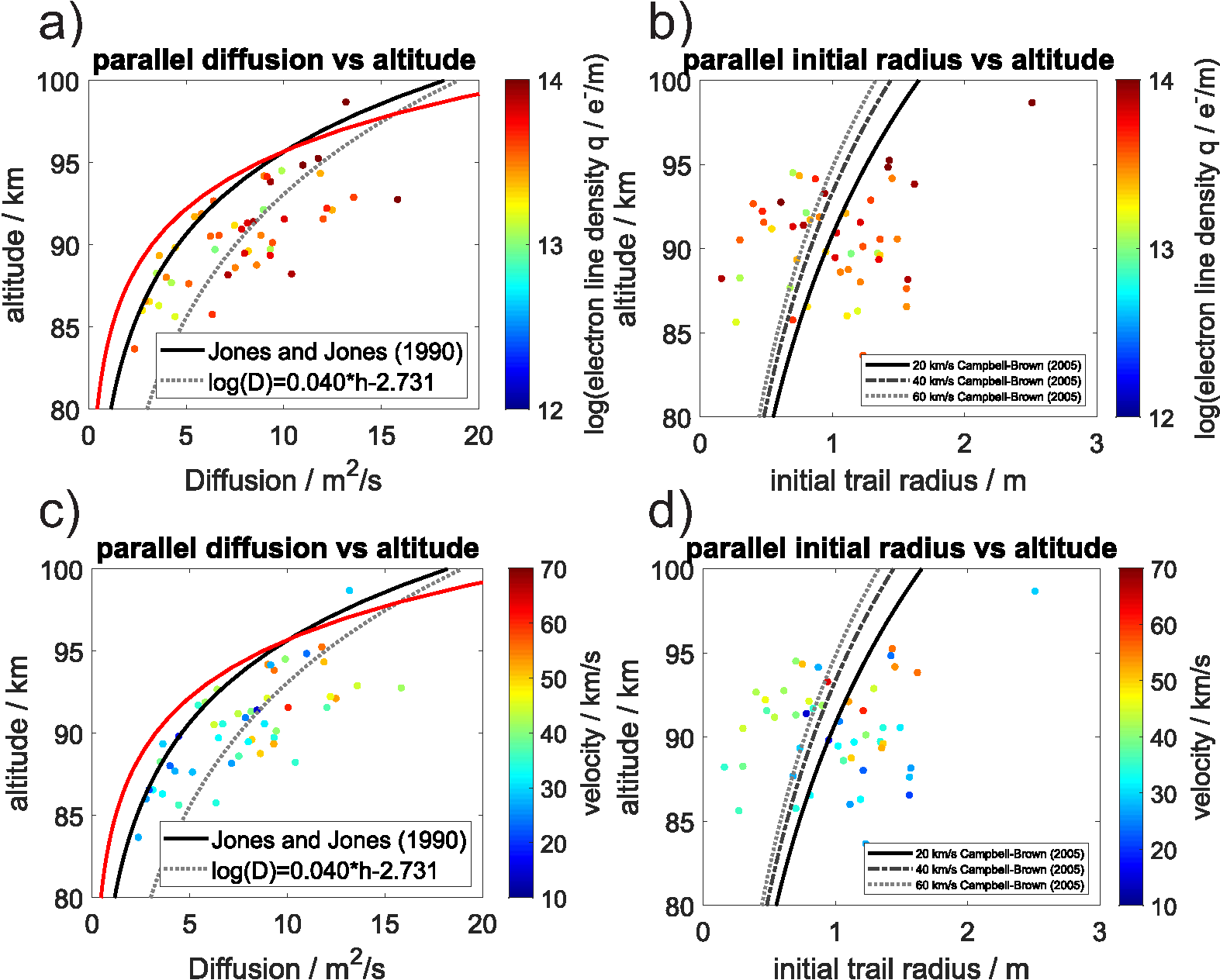}}
      \caption{Comparison of parallel scatter fits. a) Diffusion versus altitude scatter plot with color coded electron line density. b) Initial trail radius versus altitude scatter plot with color coded electron line density. c) Diffusion versus altitude scatter plot with color coded time of flight meteor velocity. d) Initial trail radius versus altitude scatter plot with color coded time of flight velocity. The initial radius as a function of altitude as measured by \citet{Jones2005a} and speed is also shown for comparison.}
         \label{full_wave_scattering_polzero}
\end{figure*}

For comparison, our selection of 50 parallel scattering meteor events are presented in Figure~\ref{full_wave_scattering_polzero}. Again panels a) and c) denote diffusion versa altitude color coded by electron line density and velocity, respectively. The initial trail radius versus altitude scatter plots are presented in panels b) and d). As predicted from the full wave scattering model, there are remarkable differences visible even with the relatively small statistics we have analyzed so far. The diffusion versus altitude plots indicate much less scattering compared to the transverse scatter meteor events and exhibit a linear correlation with altitude of 0.7 at the 90\% confidence level. The faintest meteors were observed at the lowest altitudes, but have higher electron line densities than the corresponding events for the transverse scatter case, as expected since the reflection coefficients are higher for the transverse set. The scatter plots of the initial trail radii versus altitude are characterized by values ranging from 0.25 to 1.5~m and reveal only a very weak altitude dependence, though in contrast to the transverse data set the overall fit is more similar to the results in \citet{Jones2005a}.

The scattering of the diffusion coefficients at similar altitudes of our observations might be related to the ionic composition of the diffusing meteor trail. \citet{Jones1990} presented a theoretical study of a binary system of ions in a meteor trail and how it affects the ambipolar diffusion coefficient using kinetic gas theory. Considering that the atomic or molecular mobility depends on the atomic mass of the ions, lighter meteoric ions are found farther away from the trail axis indicating a faster diffusion. However, the simulations also confirm that the electron distribution still keeps a Gaussian shape for such more complex chemical systems.

\section{Discussion}
\subsection{Magnetic field effects}
One effect that we have ignored in our modeling is the role of the Earth's magnetic field. \citet{KAISER_1969_magnetic_field} showed that there may be noticeable effects from the magnetic field on the ambipolar diffusion of meteor trails for field aligned specular meteors. They presented numerical and observational results illustrating that the ambipolar diffusion can be inhibited for meteors having velocity vectors aligned with the direction of the local field (within 1$^\circ$) and occurrence altitudes above 95~km \citep{Ceplecha1998}.\\

This effect reflects the fact that the ambipolar diffusion of a plasma trail can be affected by the external magnetic field, when the plasma becomes magnetized, which means that the ions and electrons can complete a full gyro spin around the field lines before encountering another collision. This is the case when the ratio of the gyrofrequency ($\omega_g$) to the collision frequency ($\nu$) becomes comparable. The gyrofrequency for an electron is given by;
\begin{equation}
    \omega_g=\frac{e |B|}{m_e}~~~,
\end{equation}
where $e$ is the elementary charge, $B$ is the magnetic field strength, and $m_e$ is the mass of an electron. The gyrofrequency for ions is much lower than electrons due to their higher mass. Thus, we can define a frequency ratio by
\begin{equation}
    \Omega=\frac{\omega_g}{\nu}~~.
\end{equation}

\begin{figure}
    \resizebox{\hsize}{!}
   {\includegraphics[width=14cm]{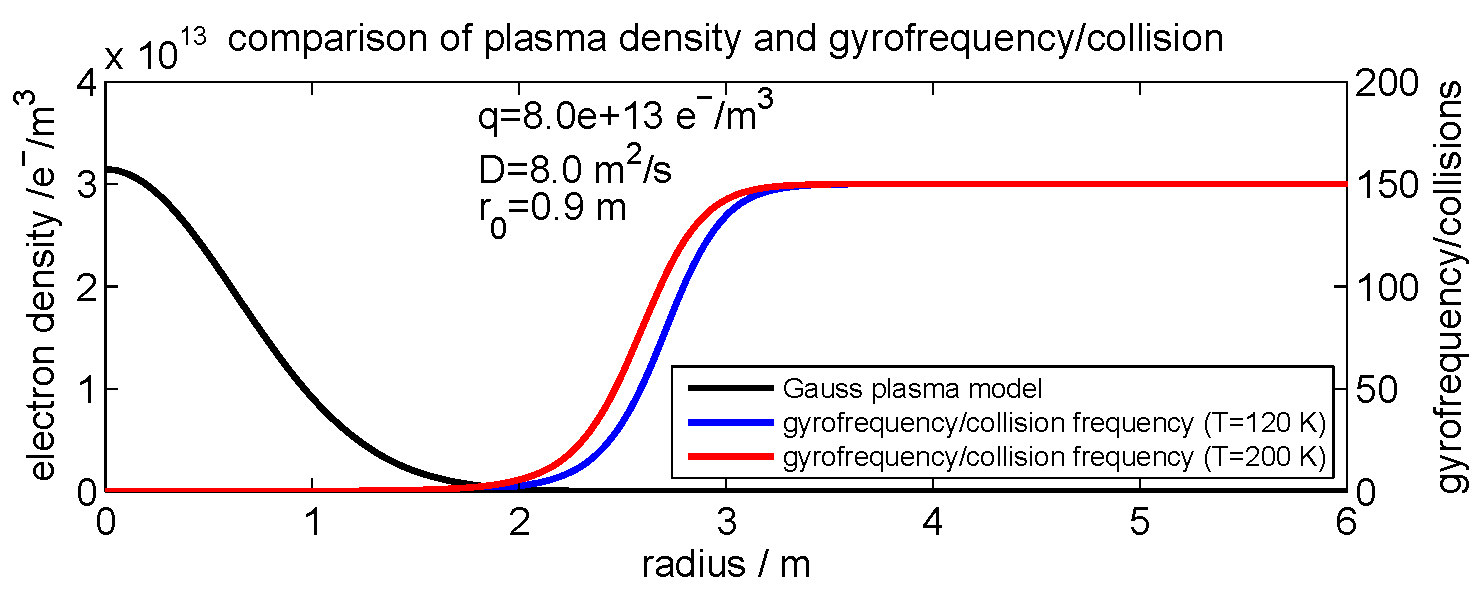}}
      \caption{Example profile for a Gaussian electron density profile (solid black line and left axis) and gyrofrequency-collision frequency ratios ($\Omega$) for summer (red) and winter (blue) (right axis).}
         \label{gyrofrequencies}
\end{figure}
In Figure~\ref{gyrofrequencies} we present an example of a radial electron density profile and the corresponding collision ratio $\Omega$. The plasma can be become magnetized for $\Omega>1$. However, for typical electron line densities detectable by CMOR ($q > 5\times10^{12}$e/m) at normal echo heights near 90 km, the peak plasma density is so high that the electron-electron, electron-ion, and electron-neutral collisions exceed by far the gyrofrequency and only when the trail is already decayed can the collision frequency ratio exhibit values larger than one.

Due to the multi-station trajectory measurements possible with CMOR, it is straight forward to compute the angle between the magnetic field lines and the meteor echo trajectories in the mesosphere/lower thermosphere. Therefore, we extracted the magnetic field data from the International Geomagnetic Reference Field (IGRF13) \citep{Thebault2015} and estimated the magnetic field strength and its inclination and declination for the location of CMOR at 90 km altitude for January 2020, essentially identical to the values expected at the time of our data collection in July 2018.

Due to the linear polarization of the CMOR radar and our selection of near transverse or parallel scattering echoes, there is only a small angular variability within our meteor data subsets. We found no meteors in our sample that had trajectories within 1$^\circ$ of the magnetic field direction, the approximate range where significant effects on the diffusion coefficient would be expected \citep{KAISER_1969_magnetic_field}. Our data show two clusters where the trajectory vectors are at angles of 20-40$^\circ$ and about 110$^\circ$ with the local magnetic field. Moreover, only one of our triple frequency echoes had a specular height above 95~km.

Based on these considerations and this analysis, we conclude that the effect of the magnetic field on the apparent diffusion coefficient is negligible for our data. \\

\subsection{Radial electron profiles}
In this study we generated numerical model estimates of the total reflection coefficient on a pulse-to-pulse basis for specular meteor echoes based on the full wave scattering methodology presented in \cite{POULTER_and_Baggaley_1977_full_wave_scat}. From this full wave scattering model we obtained reflection coefficients using three different radial plasma distributions similar to the ones presented in \cite{Marshall_and_Close_2015_FTDT}.

The Gaussian plasma distribution is obtained from the radial diffusion equation in cylindrical coordinates and has been widely adopted in prior studies \citep{Kaiser_and_Closs_1952_theoryI, Herlofson_1951}, whereas the parabolic exponential and the 1-by-r$^2$-distribution are empirically derived. \cite{Jones:1995_theory_initial_trail_radius} has expanded the theory of an initial trail radius for arbitrary radial electron distributions and showed through particle collision simulations the expected distribution is likely more centrally condensed than a Gaussian. However, he was not able at the time to repeat the full wave scattering modeling of any non-Gaussian distributions. \\

More recently, theoretical studies of meteor head echo plasma suggested a more complex and angular-dependent plasma distribution around the meteoroid. The analytical solution of \cite{Dimant:2017} assumed a dense region close to meteoroid surface, which decays radially from the center by 1/r for distances less than a mean free path length and further outside at a rate of 1/r$^2$. The analytical solution uses three essential assumptions (1) the plasma electrons follow the Boltzmann distribution, (2) most plasma ions are caused by exactly a single collision and all further collisions can be neglected and (3) the motion of the ions is not disturbed by external fields.

\cite{Sugar:2018_plasma_model_theory_and_simulation} compared the analytical results to numerical solutions and obtained similar plasma distributions. However, the numerical solution with multiple collisions showed a much better agreement with the analytical solution than the single collision simulation. Later, \cite{Sugar:2019_electro_static_simulation} conducted further particle-in-cell simulations investigating potential changes of the plasma distributions due to external electric or magnetic fields.

These theoretical works describe the plasma distribution immediately after its generation close to the meteoroid surface appropriate to timescales of order milliseconds, whereas the plasma trail observed by transverse scatter radars is much more evolved and appropriate to timescales orders of magnitude larger. On these timescales, the plasma is thermalized due to further collisions and dominated by ambipolar diffusion \citep{Ceplecha1998}. It is expected to theoretically tend toward a Gaussian distribution in time \citep{Herlofson_1951}. According to these recent numerical simulations the plasma distribution in the radial direction perpendicular to the flight path has a core region with a 1/r dependence up to one mean free path length and a 1/r$^3$ dependence for radial distances beyond a mean free path length.

We emphasize that the plasma distributions found for head echoes reflect the plasma density gradients appropriate to scales on the order of a few centimeters to a few meters and timescales of less than 1 ms at meteor echo heights (80-100 km). In contrast, our fits are appropriate to spatial scales of order kilometers and timescales of tens to hundreds of ms. Hence, the different results we find reflect changes introduced over these different timescales in the ablation evolution of the meteoroid. Our findings using model fits to the observed triple frequency CMOR observations indicate that the trail plasma for the selected events can be best described with a Gaussian plasma distribution, while a minority of cases allow fits for a 1-by-r$^2$ radial distribution. This supports the commonly held assumption of a Gaussian radial electron distribution made in past scattering studies \citep[e.g.,][]{POULTER_and_Baggaley_1977_full_wave_scat}.

However, it is important to emphasize that we could only fit a small fraction of the total echoes examined - less than 5\%. This suggests that the simple full wave scattering theory, not accounting for chemistry or fragmentation, is a poor assumption for the majority of echoes.

The initial trail radius concept was introduced with the Gaussian plasma distribution and can be generalized to other plasma distributions. The physical meaning of the initial radius is the separation between a dense core plasma region and its radial dependence and the plasma farther away from the trail axis perpendicular to the flight path. From the theoretical results of \cite{Jones:1995_theory_initial_trail_radius} and \cite{Sugar:2018_plasma_model_theory_and_simulation} the initial radius is expected to be proportional to the local atmospheric mean free path length for meteor head echoes. However, the large scatter in the initial trail radius versus altitude plots (Fig~\ref{full_wave_scattering_polninety} and Fig.~\ref{full_wave_scattering_polzero} panels b) and d), respectively) demonstrate that there is only a weak altitude dependence and no obvious connection to the mean free path. We suggest that this may reflect the role of fragmentation even for echoes, which roughly follow the full wave scattering model that lacks explicit inclusion of fragmentation effects.

Although the full wave scattering model provides an idealized theoretical description of the scattering of plane waves from cylindrical plasma columns, it approximates the observed amplitude-time profile for some echoes remarkably well. Other numerical methods used to derive the reflection coefficients using FDTD techniques, which have been applied for meteor head echoes to obtain radar cross sections \citep{Marshall_and_Close_2015_FTDT,MARSHALL_2017_CMOR}, are not yet applicable for specular backscatter interpretation due to the large domain required to simulate a full cylindrical trail of at least a few kilometers in length and several meters in radius.\\

Our quantitative results of the full wave scattering model predict a significant dependence of the echo profile on the angle between the incident plane wave and meteor trajectory. This angular dependence was confirmed by manually fitting the obtained reflection coefficients to the triple frequency data of CMOR for selected events. In particular, for the parallel scattering case, a consistent linear altitude dependence of the ambipolar diffusion is found. The manually fit reflection coefficients to the selected meteor events were consistent with Gaussian electron distributions having initial trail radii varying from 0.5 - 2 m between 85 and 95 km altitude, albeit with factor of two scatter across any given height. The initial radii for parallel scattering were systematically lower than the transverse case, likely reflecting enhanced detectability due to resonance effects. These values of the initial trail radii for parallel polarizations are comparable to those presented in \cite{Jones:1995_theory_initial_trail_radius, Jones2005a}, while the transverse scattering initial radii are systematically larger across all heights.\\

The large scatter of the diffusion versus altitude plots for the transverse scatter case might also be related to resonance effects, a result predicted by \cite{Kaiser_and_Closs_1952_theoryI}. \cite{POULTER_and_Baggaley_1977_full_wave_scat} also found resonance effects with the full wave scattering approach, but less strong. We performed several simulations to confirm our numerical results of the resonance effects.  The full wave scattering solutions represent a steady state solution corresponding to a cw-radar or a very long radar pulse. Due to the complex nature of the dielectric function, the strength of the resonance strongly depends on the collision frequencies, and thus, on the ambient atmosphere. At present all simulations are performed using the collision frequencies given by \cite{Callen_2006} and \citet{Marshall_and_Close_2015_FTDT}.\\

Atmospheric variability may also explains some part of the larger scattering in the diffusion versus altitude plot for the transverse scatter geometry, as collision frequencies will be similarly affected. The large difference that we find in apparent diffusion coefficients between parallel and transverse polarizations may be one of the causes of the large scatter commonly found in atmospheric studies of the diffusion coefficient \citep[e.g.,][]{Younger2014}, when the polarization of individual echoes is not known. Previous studies with the NAVy Global Environment Model - High Altitude (NAVGEM-HA) \citep{Eckermann:2018} on atmospheric winds showed a remarkable agreement on short timescales with meteor radar observations \citep{McCormack_2017_MLT_winds,Stober_2020_tide}. Based on this meteorological analysis, \citet{VIDA_2021_fragmentation_density} estimated the atmospheric induced neutral air density variability to be on the order of 20-30\% at mesospheric and lower thermospheric altitudes.

A potential explanation for some of the scatter we measure in the diffusion coefficients may be meteor trail chemistry, which can become complex depending on the meteor trail lifetime and height. A previous study presented by \cite{Lee_2013_ambipolar_diffusion} indicated that below 80-85 km three body electron-neutral attachments cause some deionization of the meteor trail a result confirmed by \citep{Younger2014} who also noted that ozone can affect echo durations for denser echoes at high altitudes. A more comprehensive model of the meteor trail chemistry was presented in \cite{Baggaley_1972_meteor_trail_chemistry} and \cite{BAGGALEY_1974_meteor_trail_chemistry}. These model data suggest that there are altitude- and electron-density-dependent chemical reactions, which generally become more important for timescales beyond 1 second and at altitudes below 80-85 km, which are largely outside our measurement ranges.

Furthermore, the simulations show that the background ozone concentration is an important chemical driver in limiting the lifetime of long lasting (overdense) meteors\citep{Poole1975}. These simulations also confirmed that for underdense meteors electron-ion recombination has a negligible effect \citep{BAGGALEY_1979_underdense_trail_chemistry} for the altitudes of most of our echoes.

Differential ablation due to the early release of meteoric alkali atoms may also play a minor role in modifying the diffusion coefficient, but generally occurs at altitudes above 100 km \citep{Vondrak_2008,Janches_2009_differential_ablation}, outside our height range. The diffusion of the meteor trail can also be altered in the presence of dusty particles \citep{HAVNES_2005_trail_diffusion_dust_particles} e.g., noctilucent clouds. These clouds are frequently observed at mid- and high-latitudes during the summer months, but are not present at the latitude of CMOR \citep{Thomas_2019_Sophie_NLC}.

\section{Conclusions}
Application of a full wave scattering model provides a first principles approach to describe the scattering of electromagnetic waves from meteor trails. In this study, we applied the full wave scattering model to derive quantitative reflection coefficients for specular meteor trail echoes on a pulse-to-pulse basis for comparison to measurements. Using the CMOR trajectory solution allowed estimates for speed, height and polarization angle for each specular echo. By combining this information with model reflection coefficients, we found values of $q$, D and r$_0$ to simultaneously fit synthetic echo amplitude-time profiles to actual measurements at frequencies of 17.45, 29.85 and 38.15 MHz. The full wave scattering model was used to investigate how the echo profiles also varied across three different plasma distribution models.\\

In total 50 transverse scatter and 50 parallel meteor events were manually fit. The transverse scatter events showed large variability in the diffusion versus altitude plots as well as in the initial trail radius versus altitude plots suggesting only a weak linear dependence. The large variability for transverse polarization is mostly attributed to resonance effects, which are inherent to the full wave scattering model for this geometry and strongly depend on the ambient atmospheric conditions. Ther parallel scattering geometry, in contrast, showed a linear dependence of the diffusion with altitude and less variability. Further, we confirmed that the minimum detectable electron line density for parallel polarizations is higher compared to the transverse scatter case, as expected from the full wave scattering model. Our values for initial radius between 85-95 km height varied from 0.5 - 2 m, broadly consistent with prior measurements.

We found that the majority of echoes($\approx$ 95\%) having near-parallel or near-transverse polarizations were not amenable to good fits across all three frequencies. No echoes could be fit at all frequencies for any  $q$, D and r$_0$ combination using a parabolic radial electron  density distribution. A small minority of echoes could be fit with a $1/r^2$  distribution, but the most fits were achieved using a Gaussian distribution.

The angular dependence of the scattering and the resultant echo profile further indicates that atmospheric temperature measurements obtained from the decay time can be improved by multi-frequency observations and multi-static configurations to derive trajectory information. Furthermore, our results underline that the sensitivity of a meteor radar is given by the peak power of the pulse and, thus, the corresponding field strength rather then the duty cycle. The full wave scattering model, which is presented herein, is applicable to all radars with a pulse width much longer than a wavelength including cw-meteor radars.

\begin{acknowledgements}
     This work was supported by NASA co-operative agreement 8080NSSC18M0046. PGB also acknowledges funding support from the Natural Sciences and Engineering Research council of Canada (RGPIN-2016-04433) and the Canada Research Chairs program (grant 950-231930). Calculations were performed on UBELIX (http://www.id.unibe.ch/hpc), the HPC cluster at the University of Bern.
\end{acknowledgements}



\begin{appendix} 
\section{Radar reflection coefficient tables}
Tables from full wave scattering model for 6 frequencies at 17.45, 29.85, 32.55, 36.20, 38.15, 53.5 MHz can be found in the supplement material. The tables are available as ASCII-files and have 11 columns consisting of the trail radius $r_0$ (m/s), the critical radius for the overdense to underdense transition $r_c$ (m/s), the boundary matching radius $r_b$ (m/s), the parallel and perpendicular reflection coefficients gE and gH, the phases of the parallel and perpendicular reflection coefficients, the polarization between perpendicular and parallel reflection coefficient, the electron line density q ($e^{-}/m$, the order of the Bessel and Hankel functions included  in the solution, and finally an arbitrary scaling factor for comparisons with previous studies $(kr^2)$. We compiled one look up table per plasma distribution and frequency as described in the manuscript.

\end{appendix}

\end{document}